\documentclass[aps,prd,eqsecnum,12pt,showpacs,preprintnumbers,nofootinbib]{revtex4-2}
\usepackage{geometry}                
\usepackage[pdftex]{graphicx}
\usepackage{float}
\usepackage{amssymb,amsmath,amsopn,bm,dsfont,mathrsfs}

\usepackage[mathscr]{euscript}

\usepackage[colorlinks,linkcolor=blue, pdfstartview=FitH]{hyperref}
\usepackage{enumitem}
\usepackage{txfonts}
\usepackage[compact,small]{titlesec}

\usepackage{amsbsy}

\hypersetup{linkcolor=blue,citecolor=blue,filecolor=black,urlcolor=blue}

\IfFileExists{dsfont.sty}
	{\usepackage{dsfont}
         \let\mathbb=\mathds
         \newcommand{\id}{\mathds{1}}}
	{\typeout{Package dsfont.sty was not found, using alternative macros.}
         \let\mathds=\mathbb
         \newcommand{\id}{\mbox{1 \kern-.59em \textrm{l}}}}

\renewcommand\a{\alpha}
\renewcommand\b{\beta}
\renewcommand\d{\delta}

\renewcommand\l{\lambda}
\renewcommand\r{\rho}

\newcommand\y{\upsilon}

\renewcommand\th{\theta}

\newcommand\ch{\raisebox{0.7\depth}{$\chi $}}
\newcommand\e{\epsilon}
\newcommand\g{\gamma}

\newcommand\m{\mu}
\newcommand\n{\nu}

\newcommand\p{\pi}
\newcommand\h{\eta}
\newcommand\s{\sigma}
\newcommand\f{\phi}
\newcommand\w{\omega}
\newcommand\ve{\varepsilon}
\newcommand\vk{\varkappa}

\newcommand\vf{\varphi}

\newcommand\La{\Lambda}

\newcommand\Th{\Theta}
\newcommand\D{\Delta}
\newcommand\G{\Gamma}

\newcommand\Y{\Upsilon}
\newcommand\W{\Omega}

\newcommand{\lag}{\langle}
\newcommand{\rag}{\rangle}

\newcommand{\cA}{{\cal A}}

\newcommand{\tF} {{\widetilde F}}

\newcommand{\cL}{{\cal L}}

\newcommand{\cO}{{\cal O}}

\newcommand{\cG}{{\cal G}}

\newcommand{\pa}{\partial}

\newcommand{\na}{\nabla}
\newcommand{\sdfrac}[2]{\mbox{\small$\displaystyle\frac{#1}{#2}$}}

\newcommand{\by}{{\boldsymbol \y}}

\newcommand{\GN} {G_{\scriptstyle N}}
\newcommand{\rH}{r_{\!_H}}
\newcommand{\rM}{r_{\!_M}}
\newcommand{\bea}{\begin{eqnarray}}
\newcommand{\eea}{\end{eqnarray}}
\newcommand{\be}{\begin{equation}}
\newcommand{\ee}{\end{equation}}
\newcommand{\bes}{\begin{subequations}}
\newcommand{\ees}{\end{subequations}}

\def\nbox#1#2{\vcenter{\hrule \hbox{\vrule height#2in
\kern#1in \vrule} \hrule}}
\def\sq{\,\raise0.8pt\hbox{$\nbox{.10}{.10}$}\,}
\def\sqb{\,\raise.5pt\hbox{$\overline{\nbox{.09}{.09}}$}\,}

\newcommand{\sumi}{\raisebox{-1.5ex}{$\stackrel{\textstyle\sum}{\scriptstyle i}$}}
\newcommand{\st}{{\star}}

\usepackage[T1]{fontenc} 
\usepackage[titletoc]{appendix}
\usepackage{setspace}

\allowdisplaybreaks

\begin{document}
\pagestyle{empty}

\title{The Effective Theory of Gravity and Dynamical Vacuum Energy}

\author{Emil Mottola}
\email{mottola.emil@gmail.com, emottola@unm.edu}
\affiliation{Department of Physics and Astronomy, University of New Mexico\\
Albuquerque NM 87131\\}

\begin{abstract}
  Gravity and general relativity are considered as an Effective Field Theory (EFT) at low
  energies and macroscopic distances.  The effective action of the conformal anomaly of light
  or massless quantum fields has significant effects on macroscopic scales, due to associated light
  cone singularities that are not captured by an expansion in local curvature invariants.  A compact
  local form for the Wess-Zumino effective action of the conformal anomaly and stress tensor is
  given, requiring the introduction of a new light scalar field, which it is argued should be included
  in the low energy effective action for gravity. This scalar \emph{conformalon} couples to the
  conformal part of the spacetime metric and allows the effective value of the vacuum energy,
  described as a condensate of an exact $4$-form abelian gauge field strength $F=dA$, to change 
  in space and time. This is achieved by the identification of the torsion dependent part of the
  Chern-Simons $3$-form of the Euler class with the gauge potential $A$, which enters the effective 
  action of the conformal anomaly as a $J \cdot A$ interaction analogous to electromagnetism. 
  The conserved $3$-current  $J$ describes the worldtube of $2$-surfaces that separate regions of 
  differing vacuum energy. The resulting EFT thus replaces the fixed constant $\Lambda$ of classical 
  gravity, and its apparently unnaturally large sensitivity to UV physics, with a dynamical condensate 
  whose ground state value in empty flat space is $\Lambda_{\rm eff} = 0$ identically.  By allowing 
  $\Lambda_{\rm eff}$ to vary rapidly near the $2$-surface of a black hole horizon, the proposed 
  EFT of dynamical vacuum energy provides an effective Lagrangian framework for gravitational 
  condensate stars, as the final state of complete gravitational collapse consistent with quantum 
  theory. The possible consequences of dynamical vacuum dark energy for cosmology, the cosmic 
  coincidence problem, and the role of conformal invariance for other fine tuning issues in the 
  Standard Model are discussed.

\end{abstract} 

\maketitle
\flushbottom

\tableofcontents
\pagenumbering{arabic}
\pagestyle{plain}
\vfil\break

\section{The Quantum Vacuum and Vacuum Energy in Black Holes and Cosmology}
\label{Sec:Intro}

Observations of type Ia supernovae (SN) at moderately large redshifts indicate that the expansion of the universe is
 accelerating~\cite{RiessSN:1998,PerlmutterSN:1999}. This is possible in classical general relativity (GR) only if the dominant
energy of the universe has an effective mean eq.~of state satisfying $\,\r + 3p <0$,~i.e.~assuming positive energy density
$\r >0$, it must have negative pressure. As the cosmological term $\La$ enters Einstein's eqs.~as a constant with
$p_\La = -\r_\La =  -\La/8\p \GN$ pervading all space, the SN observations taken at face value imply a $\La$ value of~\cite{DES:2019}
\vspace{-3mm}\be
\La_{\mathrm{SN}} = \W_\La \times 3\,\Bigg(\frac{H_0}{c}\Bigg)^2 \simeq
\left(\frac{\W_\La}{0.70}\right) \left(\frac{H_0}{70 {\, \rm km/sec/Mpc}}\right)^2\ \left(\frac{3.1 \times 10^{-122}  }{L_{\mathrm{Pl}}^2}\ \right)
\label{LSN}
\vspace{-3mm}\ee
when expressed in terms of the present Hubble expansion rate of $H_0 \simeq 70\,$ km/sec/Mpc, or the microscopic Planck length
$L_{\mathrm{Pl}} =   \sqrt{\hbar \GN/c^3}  \simeq 1.616 \times  10^{-33}$ cm.~respectively. Thus some $\W_\La  \simeq  70\%$
of the energy in the present universe is in the form of $\La$ dark energy, and is the principal component of the current $\La$CDM
model of cosmology.

The contrast in~(\ref{LSN}) between the dimensionless value of the cosmological term $\W_\La$, of order unity in cosmological Hubble units,
but of order $10^{-122}$ in microscopic Planck units, is striking. From the time of W.~Pauli it has been thought that $\La$ is related
to the zero-point energy density of the vacuum in quantum field theory (QFT), in which it appears as an ultraviolet (UV) divergent
sum over all field modes~\cite{WeinbergRMP:1989,RughZink:2003,AntMazEM:2007,Martin:2012}. If $\La$ is such a UV
sensitive quantity, and the short distance cutoff is of order of $L_{\mathrm{Pl}}$,  then the value of $\La_{\mathrm{SN}}$ in Planck
units represents the most severe scale hierarchy problem in all of physics, clashing with expectations of `naturalness' developed over
several decades of successful application of Effective Field Theory (EFT) methods~\cite{Leut:1994,Burg:2007,Alv:2012}.

On the other hand, if one adopts the EFT hypothesis that macroscopic gravity and the value of $\La$ at cosmological scales should be
decoupled from and not require detailed knowledge of extreme UV physics, then~(\ref{LSN}) suggests instead that the EFT of gravity is
incomplete at \emph{low energies}, and one or more additional EFT degrees of freedom are needed to account for a vacuum energy
naturally of order of the Hubble scale.

If the EFT of gravity relates $\La_{\rm eff}$ to the cosmological Hubble scale $3H_0^2$ rather than the microscopic Planck scale
$L_{\mathrm{Pl}}^{-2}$, the further implication is that $\La_{\rm eff}$ would have to become a \emph{dynamical} quantity,
i.e.~dependent upon the content and evolution of the universe, as the Hubble `constant' itself is~\cite{AntMazEM:2007}. 
Related to both possibilities of additional low energy gravitational degrees of freedom other than the metric of classical GR and 
of $\La_{\rm eff}$ becoming dynamical as a result, it is worth noting that current cosmological models already require at least 
one additional scalar (inflaton) field of unknown origin, in order to generate the present small ($\sim   10^{-5}$) CMB anisotropies 
during a very early epoch of cosmic inflation~\cite{Guthbook:1998}. This epoch is assumed to have been dominated by a 
much larger effective $\La_{\rm eff}$ vacuum energy, that is supposed to have dynamically `relaxed' to its present much smaller value.

Indications that the EFT of gravity may require some additional degree(s) of freedom relevant at macroscopic scales come also from the
quite different domain of black hole (BH) physics. EFT methods in gravity for BHs have been put into question by both the extreme
blueshifting of energy scales in the presence of horizons, invalidating the EFT assumption of decoupling of short distance from long
distance physics~\cite{Jacobsen:1993}, and by the BH `information paradox'~\cite{Preskill:1992,tHooft:1995,tHooft:2006,Mathur:2009,AMPS:2013,Giddings:2013,MottVauPT,HarlowBHs:2016}, and apparent conflict with unitary evolution it
entails~\cite{HawkUnit:1976}.
The various forms of this paradox arise from ascribing an enormous entropy to a BH, equal to $1/4$ its horizon area $A_H$ in Planck units,
\vspace{-3mm}\be
S _{\mathrm{BH}} = k_B\, \sdfrac{A_H}{4\, L_{\mathrm{Pl}}^2 } \simeq  1.1 \times 10^{77}\, k_B \, \left(\sdfrac{ M}{M_\odot}\right)^2
\label{SBH}
\vspace{-3mm}\ee
despite the assumed classical nature of the BH horizon as a mathematical causal boundary only, with no independent degrees
of freedom of its own. Significant quantum effects on the macroscopic scale of the BH horizon are also in apparent conflict with the usual EFT
approach to gravity, which relies on an expansion in local curvature invariants~\cite{Don:1994,Bur:2004,Don:2012}, since these can
yield only negligibly small corrections for large BHs with small local curvatures at their horizons. In fact, the enormous entropy~(\ref{SBH}) and
Hawking effect upon which it is predicated rely crucially upon the specification of the quantum vacuum state, which (as always in quantum
theory), requires \emph{non-local} boundary conditions, that are not determined solely by the local curvature~\cite{Boul:1975,Unruh:1976, ChrFul:1977}.

In previous work it has been noted that large quantum effects on the horizon follow quite generally from the stress tensor of the conformal
anomaly in both BH and cosmological spacetimes~\cite{EMVau:2006,GiaEM:2009,EMZak:2010}. The importance of the conformal
anomaly in the near horizon behavior of the stress tensor is a consequence of the conformal scaling behavior of the metric near the horizon
and the extreme blueshifting of local frequencies, which renders all finite mass scales irrelevant there~\cite{EMZak:2010}.
It is just this extreme blueshifting of frequencies, hence energies, that can lead to effects not accounted for in local EFT expansions based on
the assumption of decoupling and strict separation of scales. State dependent quantum vacuum entanglement and polarization effects
are contained in the effective action of the conformal anomaly, which by its nature spans multiple scales.

That the physics of BHs and vacuum energy are related is inherent also in the proposed resolution of the BH information paradox by the
formation of a gravitational vacuum condensate star with interior $\La_{\rm eff}$ eq.~of state
$p = -\r$~\cite{gravastar:2001,MazEMPNAS:2004}. This $\La_{\rm eff}$ eq.~of state in the interior of a \emph{gravastar}, collapsed
to its gravitational radius $\rM  =  2\GN M/c^2$ prevents further collapse to a BH singularity for the same reason that it causes the Hubble
expansion of the universe to accelerate, namely by defocusing (rather than the usual focusing) of worldline geodesics, avoiding the classical
singularity theorems~\cite{Penrose:1965,HawkPen:1970,HawkEllis:1973}. The localized formation of such a $p = -\r$ gravitational
vacuum condensate within an ultra compact star can occur only if there is at least one additional degree of freedom in the low energy EFT of
gravity, whose variation allows $\La_{\rm eff}$ to change abruptly at or near $r = \rM$. This turns the BH horizon from a mathematical surface
to a physical phase boundary layer with a positive surface tension~\cite{MazEM:2015,BelGonEM1:2022,BelGonEM2:2022}.

The purpose of this paper is to propose and develop the EFT of low energy gravity, deduced from general principles of QFT in curved space
and the conformal anomaly, in which finite dynamical vacuum energy is consistently described as a scalar vacuum condensate.
The description of vacuum energy by the scalar dual to an exact $4$-form abelian field strength $F=dA$ requires $\La_{\rm eff} \ge  0$ 
with $\La_{\rm eff} =  0$ the unique value of lowest energy in flat space, independently of UV physics. When the $3$-form potential
$A$ is identified with the Chern-Simons $3$-form of the Euler class, a $J \cdot A$ interaction is induced by the conformal anomaly 
effective action, in analogy with electromagnetism. The $3$-current $J$ source for $F$ describes the worldtube of $2$-surfaces 
that separate regions of differing vacuum energy $\La_{\rm eff}$, which therefore becomes spacetime dependent. The observational 
implications of this EFT extension of classical GR and identification of the relevant low energy degrees of freedom describing $\La_{\rm eff}$ 
as a dynamical gravitational vacuum condensate for both BH physics and cosmological vacuum dark energy can then be studied in detail.

The metric and curvature conventions of the paper are those of MTW~\cite{MTW}, while the definitions and conventions for tetrads
and differential forms used are reviewed in appendix~\ref{App:Tetrad}. A second appendix~\ref{App:Susc} is devoted
to the topological aspects of the Euler density, associated Chern-Simons $3$-form and physical interpretation of the new
constant $\vk$ introduced in the EFT, as a torsional topological susceptibility of the gravitational vacuum.

\section{Relevance of the Conformal Anomaly to Macroscopic Gravity}
\label{Sec:Macro}

If one takes as the basic building block of a gravitational theory the spacetime metric $g_{\m\n}(x)$, with the requirements that the field
eqs.~must transform as tensor eqs.~under general coordinate transformations, and be no higher than second order in derivatives
of the metric, one arrives at the classical theory of general relativity (GR). This is described by the classical action
\vspace{-3mm}\be
S\!_{cl} =S_{ \rm EH} - \sdfrac{\La}{8 \p \GN}\int\! d^4 x \sqrt{-g} = \sdfrac{1}{16\p \GN} \int\! d^4x \sqrt{-g}\,\big(R -2 \La \big)
\label{Scl}
\vspace{-4mm}\ee
namely the Einstein-Hilbert (EH) action involving the Ricci curvature scalar $R$, second order in derivatives of the metric,
or first order in derivatives of the symmetric Christoffel connection
\vspace{-3mm}\be
\G^\l_{\ \,\m\n} = \G^\l_{\ \,\n\m}= \sdfrac{1}{2} g^{\l\r}\,\Big(\! -\pa_\r g_{\m\n} + \pa_\m g_{\n\r} + \pa_\n g_{\m\r}\Big) \,,
\label{Chris}
\vspace{-3mm}\ee
together with the cosmological constant term $\La$, involving no metric derivatives. In modern terms the requirement
of an action $S\!_{ cl}$ composed of sums of all integrals of scalars that are invariant under general coordinate
transformations (up to possible surface terms) and yield field eqs.~no higher than second derivatives of the metric is
just what is meant by a low energy EFT of gravity, since local invariant terms higher order in derivatives of the metric
are negligible at low energies or long wavelengths. In~(\ref{Scl}) the constants $\GN$ and $\La$ can be determined
only by experiment or astronomical observations. At this purely classical level, if $\La$ is given by observations to be~(\ref{LSN})
there is no naturalness problem, for there is no other scale in the classical theory to which it can be compared.

As \'E.~Cartan pointed out soon after the appearance of GR~\cite{Cartan:1922,Hehl:1976}, the most general setting of an 
affine geometry and differential manifold allows also for a non-zero torsion, which is described by an anti-symmetric part of the connection
$\G^\l_{\ \,[\m\n]} \neq 0$ that Einstein had assumed to be vanishing, as the simplest realization of the Equivalence Principle.
An anti-symmetric part of $\G^\l_{\ \,\m\n}$ drops out of the geodesic eq.~for the worldlines of freely falling point particles in any case.
Einstein-Cartan theory allows the connection $\G^\l_{\ \,\m\n}$ and functions of it to be treated as dynamical variables in their own 
right, {\it a priori} independent of the spacetime metric~\cite{Hehl:1976}, a property that will be exploited in Sec.~\ref{Sec:CS}.

Let us note that the problems of reconciling classical GR with QFT first appear with the stress-energy tensor $T^{\m\n}$, which is
treated as a completely classical source in Einstein's eqs., whereas $\hat T^{\m\n}$ is a UV divergent operator in QFT. Since matter and 
radiation in the Standard Model (SM) are certainly quantum in nature, replacing the quantum operator $\hat T^{\m\n}$ by its 
renormalized expectation value $\lag \hat T^{\m\n}\rag$ in a semi-classical approximation amounts to two different assumptions, 
which should be recognized and distinguished at the outset.

The first assumption is that UV divergences of QFT are to be removed by counterterms involving up to dimension-four curvature invariants,
such as $R_{\a\b\m\n}R^{\a\b\m\n}, R_{\a\b}R^{\a\b}, R^2$, by adding them to the effective action with finite renormalized coefficients,
leaving the low energy EFT unchanged. These terms $\sim (\pa^2 g_{\m\n})^2$, involve up to four derivatives of the metric, in contrast to
the dimension-two EH action.  Since quantum theory introduces a new scale $L_{\mathrm{Pl}}$, this standard renormalization procedure
amounts to the assumption that such higher derivative terms may be important only on the corresponding Planck energy
scale of $M_{\mathrm{Pl}} c^2 =1.221 \times 10^{19}$\,GeV.  Since this scale is so much higher than those generally encountered either
in terrestrial accelerators or astrophysics, the reasonable assumption of an EFT approach is that Planck scale physics decouples
from the low energy EFT, so that knowledge of the UV completion or full quantum theory is not needed to describe macroscopic gravitation.

Less often noted is a second critical assumption in the replacement of the stress tensor source in Einstein's eqs.~by a finite renormalized
$\lag \hat T^{\m\n}\rag$, treated classically, namely that the operator $\hat T^{\m\n}$ can be well approximated by its sharply peaked
mean value. It is easily verified in QFT that even after renormalization there are quantum fluctuations from the mean, and
e.g.~$\big\lag \hat T^{\a\b} (x)\, \hat T^{\m\n}(y)\big\rag - \big\lag \hat T^{\a\b}(x) \big\rag\, \big\lag \hat T^{\m\n}(y)\big\rag \neq 0$
at one-loop order~\cite{AndMolEM:2003}. Connected higher point correlation functions of this kind probe the polarization and entanglement
properties of the quantum vacuum even at macroscopic scales. These quantum correlators exhibit operator product singularities as
$x \to y$. All is well with these UV singularities since dimensional analysis, as well as explicit calculations show that these quantum
correlations grow large as $L_{\mathrm{Pl}}^2/\ell^2$ relative to $\lag \hat T^{\m\n}\rag$ only for metric variations on the length
scale $\ell \lesssim L_{\mathrm{Pl}}$.
Neglecting these short distance correlations at scales $\ell \gg L_{\mathrm{Pl}}$ reduces then to the first decoupling assumption of EFT.

There is however a second kinematic regime where the Lorentz invariant distance $(x-y)^2 \to 0$, even for $x\neq y$, and where
two and higher point quantum correlation functions of $\hat T^{\m\n}$ can become large, namely on the light cone. Since light cones
extend over arbitrarily large distances, lightlike correlations are \emph{not} limited to the ultrashort $L_{\mathrm{Pl}}$, but can
lead to \emph{macroscopic} quantum effects, in particular on null horizons~\cite{EMVau:2006}, which can be relevant in both 
BH and cosmological spacetimes with positive $\La$, such as de Sitter space. The two very different sorts of quantum effects, short 
distance UV vs.\ macroscopic lightlike correlations are distinct and require two quite different EFT treatments.

The short distance UV quantum corrections to GR are taken into account by adding to the action~(\ref{Scl}) of classical GR
the expansion in ascending powers of higher derivatives of local invariants, divided by appropriate powers of the UV cutoff scale,
expected to be the Planck scale $M_{\mathrm{Pl}}$ for gravity. This is the most common EFT
approach~\cite{Don:1994,Bur:2004,Don:2012}. As already mentioned, it amounts to the decoupling assumption 
common to all EFT approaches, based on the decoupling theorem of massive states in the UV from the low energy degrees 
of freedom~\cite{AppCar:1975}.

On the other hand it has also been known for some time that QFT anomalies are not captured by such an expansion in higher
order local invariants, nor are they suppressed by any UV scale. Anomalies are associated instead with the fluctuations of massless fields
which do not decouple, and which lead to $1/k^2$ poles in momentum space correlation functions, that grow large on the light cone
$k^2 \to 0$ rather than the extreme UV regime $k^2 \sim  M_{\mathrm{Pl}}^2$. The prototype of this light cone pole is the Schwinger
model of $1+1$ dimensional massless electrodynamics and its chiral anomaly~\cite{Schw:1962}, which extends to the two-dimensional
conformal anomaly in curved space~\cite{Poly:1981,BlaCabEM:2014}. In $3+1$ dimensional flat space, light cone poles are found in
explicit calculations in the triangle anomaly diagrams of $\lag  \hat J_5^\l \hat J^\a \hat J^\b\rag, \lag \hat T^{\m\n} \hat J^\a \hat J^\b\rag$
in massless QED$_4$~\cite{GiaEM:2009,ArmCorRose:2009}, and in the stress tensor three-point correlator
$\lag  \hat T^{\a\b} \hat T^{\g\l}  \hat T^{\m\n}\rag$ of a general conformal field theory (CFT),  by solution of the conformal Ward Identities
in momentum space~\cite{BzoMcFSken:2018,TTTCFT:2019}.

Quite contrary to the decoupling hypothesis, quantum anomalies lead instead to the principle of \emph{anomaly matching} from UV
to low energy EFT~\cite{tHooft:1979}. In the strong interactions, the chiral anomaly of the UV theory, QCD, survives to low energies,
requiring a specific Wess-Zumino (WZ) addition to the low energy meson EFT~\cite{WZ:1971,Leut:1994}, which is not suppressed
by any high energy scale, and is technically a marginally relevant operator in the infrared (IR). Indeed, as befits being associated with
light cone singularities, the chiral anomaly has both UV and IR features. The successful prediction of the low energy $\p^0 \to 2\g$ decay
rate provides a window into the UV and evidence for the $S\! U(3)^{\rm color}$ group and fractional charge assignments of quarks that
helped establish QCD as the UV theory of the strong interactions~\cite{FritGellLeut:1973,TreiJackGross:2015,Bertlbook}.

For gravitational theory it is the conformal anomaly in the trace of the stress-energy tensor $\hat T^{\m\n}$ that is
associated with $1/k^2$ anomaly poles in higher point correlation functions, such as $\lag \hat  T^{\a\b} \hat  T^{\g\l}  \hat T^{\m\n}\rag$.
Such light cone singularities imply the existence of at least one additional light (\emph{a priori} massless) degree of freedom 
in the low-energy EFT of macroscopic gravity, that is not accounted for in the classical action of GR~(\ref{Scl}), nor by
an expansion in higher order local curvature invariants.

A representative of the light cone singularities and massless pole associated with conformal anomalies is afforded by the
effective action of two-dimensional gravity coupled to conformal matter~\cite{Poly:1981,Poly:1987}
\vspace{-4mm}\be
S_{ \rm anom}^{\rm NL,\, 2D}[g]= - \frac{c_m}{96\p}\int d^2 x \sqrt{-g(x)}  \int  d^2 y \sqrt{-g(y)} \ R(x)\, \big( \sq^{-1}\big)_{xy} \, R(y)
\label{Poly2D}
\vspace{-1mm}\ee
where $c_m = N_s + N_f$ is the central charge, given by the sum of the numbers of massless scalar and fermion fields, and
$\big( \sq^{-1}\big)_{xy}$ denotes the Green's function inverse of the scalar wave operator. This exhibits the light cone pole of the 2D
conformal anomaly, appearing already in the connected two-point function $\lag \hat T^{\a\b}(x) \hat T^{\m\n}(y)\rag$ of the underlying
CFT. The massless scalar pole in~(\ref{Poly2D}) indicates that there is an additional scalar degree of freedom in 2D gravity coupled
to conformal matter. 

The scalar degree of freedom can be made explicit by expressing the non-local anomaly effective action~(\ref{Poly2D})
in the equivalent local form
\vspace{-4mm}\be
 S_{\cA}^{\rm 2D}[g;\vf]= - \frac{c_m}{96\p}\int d^2 x \sqrt{-g}\, \Big\{g^{\m\n} \big(\na\!_\m \vf\big )\big(\na\!_\n \vf\big)- 2R \vf\Big\}
\label{2Danom}
\vspace{-3mm}\ee
by the introduction of the scalar field $\vf$ describing a collective spin-0 degree of freedom, which is linearly coupled to $R$, and
whose massless propagator gives rise to the light cone singularities of the underlying massless CFT~\cite{BlaCabEM:2014}.
Variation of~(\ref{2Danom}) with respect to $\vf$ gives its eq.~of motion $-\sq \vf  =  R$, which when solved for $\vf =  - \sq^{-1} R$
and substituted back into~(\ref{2Danom}) returns the non-local form of the effective action~(\ref{Poly2D}). At the same time variation
of~(\ref{2Danom}) with respect to the metric $g_{\m\n}$ yields the stress tensor $T^{\m\n}_\cA [g;\vf]$ whose trace is
$-(c_m/24 \p)\sq \vf  =  c_m R /24 \p$ which is the 2D conformal anomaly. Taking two variations with respect to $g_{\a\b}(x)$
and $g_{\m\n}(y)$ yields the connected two-point CFT correlator $\lag \hat T^{\a\b}(x) \hat T^{\m\n}(y)\rag$ which
exhibits a massless $1/k^2$ light cone pole in flat space~\cite{BlaCabEM:2014}. Clearly such a massless scalar degree of freedom
affects the macroscopic behavior of 2D gravity~\cite{KPZ:1988}, but cannot be described by a local action in curvature invariants alone,
as is the very nature of an~anomaly.

\section{The Effective Action of the Conformal Anomaly and Conformalon Scalar}
\label{Sec:Anom}

The anomalous Ward identities for all higher point quantum correlation functions of the stress tensor containing anomalous light cone
singularities can be derived by functional variation of the basic one-point expectation value of $\lag \hat T^{\m\n}\rag$ in a general
curved space background~\cite{TTTCFT:2019}. In $D = 4$ this mean value, defined and renormalized by any method that preserves
its covariant conservation, results in it acquiring an anomalous trace in background gravitational and gauge fields, the general form of
which is~\cite{CapperDuff:1974,Duff:1977,BirDav}
\vspace{-3mm}\be
\big\lag \hat T^\m_{\ \ \m} \big\rag \equiv g_{\m\n}\, \big\lag \hat T^{\m\n} \big\rag
= b\, C^2 + b'\, \left( E - \tfrac{2}{3} \sq R\right) + \sumi \ \b_i\, \cL_i
\label{tranom}
\vspace{-2mm}\ee
even if the underlying QFT is conformally invariant at tree level, and one might have expected this trace to vanish. In~(\ref{tranom})
\vspace{-3mm}\be
E =R_{\a\b\g\l}R^{\a\b\g\l} - 4 R_{\a\b}R^{\a\b}  + R^2\,,\qquad
C^2 = R_{\a\b\g\l}R^{\a\b\g\l} - 2 R_{\a\b}R^{\a\b}  + \sdfrac{1}{3} R^2
\label{ECdef}
\vspace{-3mm}\ee
are the Euler-Gauss-Bonnet invariant and the square of the Weyl conformal tensor respectively.  The $b, b', \b_i$ coeffcients
in~(\ref{tranom}) are finite dimenionless coefficients (in units of $\hbar$) that depend only upon the number and spin of the
massless conformal fields contributing to the anomaly, i.e.~
\vspace{-3mm}\be
b = \sdfrac{1}{(4 \pi)^2 }\,\sdfrac{1}{120}\, \Big(N_s + 6 N_f + 12 N_v\Big)\,,
\qquad b'= -\sdfrac{1}{(4 \pi)^2 }\, \sdfrac{1}{360}\,\Big(N_s + 11 N_f + 62 N_v\Big)
\label{bbprime}
\vspace{-2mm}\ee
where $(N_s, N_f, N_v)$ are the number of free conformal scalar, Dirac fermion, and gauge vector fields respectively. Interactions
of the massless or light QFT degrees of freedom are taken into account by the ${\cal L}_i$ terms in~(\ref{tranom}), which denote
invariant Lagrangians to which these fields are coupled, such as $\cL_F = F_{\a\b}F^{\a\b}$ for light charged
particles coupled to electromagnetism, or $\cL_G  = {\rm tr}\, \big\{G_{\a\b}G^{\a\b}\big\}$ for light quarks coupled to the
$S\! U(3)^{\rm color}$ gluonic gauge fields of QCD. The $\b_i$ are proportional to the $\b$-functions of these couplings.

All terms in~(\ref{tranom}) are dependent only upon the low energy QFT particle content, independently of a UV cutoff or the Planck scale.
They are therefore independent of UV physics and with fixed matter content cannot be removed in any metric theory of gravity with a
covariantly conserved stress tensor. Successive variations of~(\ref{tranom}) with respect to the arbitrary metric background yield the
anomalous trace Ward identities that the stress tensor correlators must satisfy, even in flat space. Note that these variations are independent
of the purely local counterterms needed to define a renormalized $\big\lag \hat T^\m_{\ \,\m} \big\rag$.

To construct the effective action corresponding to~(\ref{tranom}) one notes that upon multiplying~(\ref{tranom}) by $\sqrt{-g}$,
the various terms transform as
\vspace{-6mm}
\begin{subequations}
\begin{align}
\sqrt{-g}\,C^2 & \to  \sqrt{-g}\,C^2\,\label{Csig}\\
\sqrt{-g}\,\cL_i & \to  \sqrt{-g}\,\cL_i \label{Lsig}\\
\sqrt{-g}\,\left(E - \tfrac{2}{3}\sq R\right) & \to  \sqrt{- g}\,
\left(E - \tfrac{2}{3}\sq R\right) + 4\,\sqrt{-g}\,\D_4\,\s  \label{Esig}\
\end{align}\label{CEsig}\end{subequations}
\vspace{-1.3cm}

\noindent
under the conformal variation of the metric $g_{\m\n} \to  e^{2\s}g_{\m\n}$. Here
\vspace{-3mm}\be
\D_4 \equiv \na_\m \left(\na^\m\na^\n +2R^{\m\n} - \tfrac{2}{3} R g^{\m\n} \right)\na_\n
=\sq^2 + 2 R^{\m\n}\na_\m\na_\n - \tfrac{2}{3} R \sq + \tfrac{1}{3} (\na^\m R)\na_\n
\label{Deldef}
\vspace{-3mm}\ee
is the (unique) fourth order scalar differential operator that is conformally covariant~\cite{Panietz,Rieg:1984}
\vspace{-4.5mm}\be
\sqrt{-g}\, \D_4 \to \sqrt{-g}\,\D_4
\label{invfour}
\vspace{-4.5mm}\ee
for arbitrary $\s(x)$. It is thus the four-dimensional analog of the second order wave operator $\sq$ which has
the conformal property $\sqrt{-g}\, \sq \to \sqrt{-g}\,\sq$ in $D = 2$.

Using these properties under conformal variations and the fact that the anomaly density
\vspace{-3mm}\be
\cA  \equiv \sqrt{-g}\,\big\lag \hat T^\m_{\ \ \m} \big\rag = \frac{\d  S _{\rm anom}} {\d \s}
= 2\, g_{\m\n}\,  \frac{\d  S _{\rm anom}} {\d g_{\m\n}}
\label{varS}
\vspace{-3mm}\ee
is the conformal variation of an effective action of the anomaly, it is straightforward to find an action satisfying~(\ref{varS}). The non-local form of this anomaly effective action is~\cite{Rieg:1984,AntEM:1992,AntMazEM:1992,EMSGW:2017}
\vspace{-3mm}
\begin{align}
S^{\mathrm{NL}}_{  \rm anom}[g] & =\sdfrac {1}{4} \int  d^4 x\sqrt{-g_x}\, \left(E - \tfrac{2}{3}\sq R\right)_{ x}
 \nonumber \\
&\quad \times \int  d^4 y\sqrt{-g_y}\,\big(\D_4^{-1}\big)_{xy}\,\bigg\{\sdfrac{b'}{  2} \left(E - \tfrac{2}{3}\sq R\right)
+  b\,C^2 + \sumi\, \b_i\,\cL_i\bigg\}_y
\label{Snonl}
\end{align}
\vspace{-1.2cm}

\noindent
where $\big(\D_4^{-1}\big)_{xy}$ denotes the Green's function inverse of the fourth order differential operator~(\ref{Deldef}) between the spacetime points $x$ and $y$, indicated by subscripts in~(\ref{Snonl}), so that
$\int  d^4 x \, \big(  \sqrt{-g}\D_4\big)_x \big(\D_4^{-1}\big)_{xy} = 1$, for all $y$. The non-local effective
action~(\ref{Snonl}) of the conformal anomaly in $D = 4$ is the analog of non-local effective action~(\ref{Poly2D}) in $D = 2$ spacetime dimensions. Note that due to the appearance of the fourth order curvature invariants
in~(\ref{tranom}) and~(\ref{Snonl}) of $D = 4$, a single light cone $1/k^2$ pole of the anomaly first appears for \emph{three}
variations of~(\ref{Snonl}) with respect to the metric, in the $3$-point correlation function
$\lag \hat T^{\a\b}\hat T^{\g\l}\hat T^{\m\n}\rag$ in flat space. This has been checked explicitly for a general
CFT in~\cite{TTTCFT:2019}.

The anomaly effective action~(\ref{Snonl}) is one term in the full one-particle irreducible (1PI) effective action obtained
by integrating out all the matter/radiation fields in a fixed but arbitrary metric and background gauge fields. It is possible
to classify all the terms in the full 1PI effective action of QFT into three general classes, so that the full 1PI action may be
expressed as the sum~\cite{MazEMWeyl:2001}
\vspace{-4mm}\be
S\!_{\rm 1PI}[g] = S\!_{\rm local}[g] + S^{\!\mathrm{NL}}_{ \!\rm anom}[g] + S\!_{\rm inv}[g]
\label{S1PI}
\vspace{-4mm}\ee
of (i) purely local, (ii) non-local anomalous, and (iii) non-local invariant under the action of the local Weyl transformation
$g_{\m\n} \to  e^{2\s}g_{\m\n}$. The classification of terms~(\ref{S1PI}) for the possible quantum
corrections to the effective action for gravity and the conformal anomaly effective action~(\ref{Snonl}) applies in a general curved space
background, and hence is considerably more general than expansions around flat space~\cite{BarvVilk:1985,BarvVilk:1990,BuchOdinShap}.
The local terms $S\!_{\rm local}$ are the ones usually considered in EFT approaches including the clearly IR relevant cosmological constant
and EH term of~(\ref{Scl}) which scale as $e^{4\s_0}$ and $e^{2\s_0}$ under global Weyl rescalings respectively, together with terms
higher order in local curvature invariants, divided by some high energy UV energy scale, which scale as $e^{-2 n \s_0}$ for $n  \ge  0$.
For $n  > 0$ these local terms are irrelevant in the IR. The $n = 0$ terms are marginal and require special care.

The local $R^2$ and $C^2$ actions are neutral under global Weyl rescalings, while $S^{\!\mathrm{NL}}_{\! \rm anom}$
of~(\ref{Snonl}) is unique (up to the $b,b', \b_i$ coefficients and $S _{\rm inv}$) in scaling \emph{linearly} with $\s_0$, i.e.~logarithmically
under the global rescalings of the metric and distance scales. Hence the anomaly terms~(\ref{Snonl}) can grow to importance in the IR
and are classified as marginally relevant. All other $n = 0$ terms satisfying $S\!_{\rm inv}[e^{2\s} g]  = S\!_{\rm inv}[g]$
are neutral under all Weyl rescalings, and hence do not grow in the IR. While contributions to low energy gravity of the Weyl invariant
terms $S\!_{\rm inv}[g]$ cannot be excluded, the minimal assumption is to take only the relevant $n < 0$ local terms of~(\ref{Scl}) and
the logarithmic non-local~(\ref{Snonl}) as the basis for an EFT treatment of gravity. One could add to~(\ref{tranom}) a $\sq R$ term
with an arbitrary coefficient, but since this term is the trace of the metric variation of a local $R^2$ action, it is classified with the local
terms $S\!_{\rm local}$ and not as part of the true anomaly, which is \emph{not} the trace of a variation of any local action~\cite{MazEMWeyl:2001}.

Like the chiral anomaly in QCD, the conformal anomaly is intrinsically a non-local quantum effect and like~(\ref{Poly2D}),
$S^{\!\mathrm{NL}}_{ \!\rm anom}[g]$ is non-local in terms of the original metric and curvature variables. The non-local 1PI effective action
of the conformal anomaly~(\ref{tranom}) in 4D can nevertheless be rendered in a compact local
form~\cite{Rieg:1984,ShaJac:1994,EMZak:2010,EMSGW:2017} by introducing an additional local scalar field $\vf$ and
making the replacement
\vspace{-1cm}
\begin{align}
\hspace{2cm}&S ^{\mathrm{NL}}_{\rm anom}[g] \to  S\! _\cA[g; \vf]=\nonumber\\[0.5ex]
& \hspace{-2cm}\sdfrac{b' }{2\,} \int  d^4 x\sqrt{-g}\,\bigg\{\! - \left(\sq \vf\right)^2
+ 2\,\Big(R^{\m\n}  -  \tfrac{1}{3} Rg^{\m\n}\Big) \,(\na_\m\vf)\,(\na_\n\vf)\bigg\}  
+ \sdfrac{1}{2}  \int  d^4 x\,\cA\,\vf  \,.
\label{Sanom}
\end{align}
\vspace{-1.2cm}

\noindent
Since $S\!_\cA[g; \vf]$ is quadratic in $\vf$ with kinetic term $-(b'/2) \int d^4 x\sqrt{-g}\, \vf \D_4 \vf$, variation of~(\ref{Sanom}) 
with respect to $\vf$ gives the linear eq.~$b'  \sqrt{-g}\,\D_4 \vf = \cA/2$, which when solved for $\vf$ and substituted back into
(\ref{Sanom}) returns the non-local form of the effective action~(\ref{Snonl}), up to a surface term and a Weyl invariant term that 
can be absorbed into $S\!_{\rm inv}[g]$. The anomaly effective action~(\ref{Sanom}) is the 4D analog of (\ref{2Danom}), and both 
are non-trivial solutions of the WZ consistency condition \cite{AntMazEM:1992, MazEMWeyl:2001, EMSGW:2017}
\vspace{-4mm}\be
S\!_\cA[e^{-2 \s} g; \vf] = S\!_\cA[g; \vf + 2 \s] - S\!_\cA[g; 2 \s]
\label{SanomWZ}
\vspace{-4mm}\ee
under Weyl conformal transformations. The scalar degree of freedom $\vf$ is therefore closely related (by the shift $\vf \to \vf + 2\s$)
to the conformal factor of the spacetime metric to which it couples.

The general classification~(\ref{S1PI}) does not preclude the possibility that $S _{\rm inv}$ might contain additional light cone singular
terms relevant in the IR, implying additional low energy degrees of freedom besides $\vf$. However only a single scalar $\vf$ is necessary
to account for the known anomaly. Although it is possible to introduce additional low energy degrees of freedom, as was done
in~\cite{ShaJac:1994,EMVau:2006},  there is no general symmetry-based reason to do so, and the simplest possibility is the minimal
one of $S\!_\cA [g;\vf]$, with $\vf$ the single scalar carrying all the conformal transformation properties of the WZ effective action
and identity~(\ref{SanomWZ}). Thus~(\ref{Sanom}) is the minimal addition to the classical action~(\ref{Scl}) needed to
take into account the light cone singularities of correlation functions of the stress tensor of quantum matter.
Also although $\vf$ resembles the dilaton of string theory in some respects, it should be distinguished from it,
because of the specific WZ identity~(\ref{SanomWZ}) required by the Weyl cohomology of the anomaly~\cite{MazEMWeyl:2001},
which reflects its quite different physical origin. The scalar $\vf$ is a \emph{collective} mode, composed of a quantum correlated pair
of massless SM fields contributing to $\lag \hat T^{\m\n}\rag$~\cite{GiaEM:2009, EMZak:2010}, similar to the Schwinger boson
in QED$_2$, or a $U(1)^{\rm ch}$ flavor singlet ($\h$ or $\h'$) meson in low energy QCD. For this reason the distinct term of
\emph{conformalon} is reserved for $\vf$.

As an explicit solution of the WZ consistency condition~(\ref{SanomWZ}), the effective action~(\ref{Sanom}) is a generating
functional that reproduces all the anomalous conformal Ward identities of a CFT. Since $\vf$ has zero scaling dimension, and
its propagator $\big(\D_4^{-1}\big)_{xy}  \sim  \log(x-y)^2$ in position space,~(\ref{Sanom}) scales logarithmically
under global Weyl rescalings, and is a marginally IR relevant operator under finite size renormalization group scaling~\cite{Wilson:1975}.
In $S\!_\cA[g; \vf]$ the $b, b'$ coefficients are to be fixed by experiment, as they may receive contributions from
the gravitational field(s) themselves~\cite{AntMazEM:1992}. These coefficients may become dependent upon
the energy scale as well, since the number and spin of fields that can be considered light enough to be effectively
massless, and contributing to the conformal anomaly, is scale dependent.

As in 2D the WZ effective action of the anomaly $S\!_\cA$ is not purely a local functional of higher order curvature
invariants, unlike the higher dimensional quantum corrections to GR usually considered~\cite{Stelle:1977, Don:1994,Don:2012}.
Being derived from non-local quantum fluctuations of massless fields, the effects of~(\ref{Sanom}) need not be negligible
on macroscopic scales far greater than $L_{\mathrm{Pl}}$, and generally are relevant on null horizons, even when local curvatures are
small there~\cite{EMVau:2006,EMZak:2010,ThomUrbZhit:2009}.  The massless excitations described by $\vf$ do not generally
decouple, so they can have physically relevant effects at low energies, which must be studied on a case by case basis, and
particularly in cases where naive EFT decoupling arguments would seem to fail.  In fact the covariantly conserved stress tensor
\vspace{-4mm}\be
T\!_\cA^{\ \m\n}[g;\vf] \equiv \sdfrac{2\!}{\!\!\sqrt{-g}} \, \sdfrac{\d}{\d g_{\m\n}} \ S\!_\cA[g;\vf]
\label{anomT}
\vspace{-4mm}\ee
derived from~(\ref{Sanom}), whose trace is~(\ref{tranom}), generally grows without bound like $(r-\rM)^{-2}$, resp. $(r-\rH)^{-2}$,
as either the Schwarzschild or de Sitter horizons at $\rM$ or $\rH$ are approached~\cite{EMVau:2006},
cf.~section~\ref{Sec:GravCond}.

The effective action $S\!_\cA$ thus amounts to a specific addition to Einstein's GR,  consistent with, and in fact required by first
principles of QFT, general covariance, and the general form of the conformal anomaly~(\ref{tranom}). It is a relevant addition in both the
mathematical and physical sense~\cite{MazEMWeyl:2001}, capturing the macroscopic light cone singularities of anomalous
correlation functions. It is therefore a necessary part of the low energy EFT of gravity, and should be added to~(\ref{Scl}) of classical GR,
much as the WZ term must be added to the low energy meson theory to account for the chiral anomaly of QCD.

\section{The Cosmological Term as a $4$-Form Gauge Field}
\label{Sec:F}

The  action~(\ref{Sanom}) of the conformal anomaly is the first essential element in the EFT of gravity taking macroscopic quantum
effects into account. Quite apart from and independent of the anomaly, there is a second element that plays an essential role in
the characterization of vacuum energy and resolution of the naturalness problem of $\La$. This is the observation that the constant $\La$ term
in~(\ref{Scl}) can be reformulated in terms of an abelian gauge theory as follows~\cite{Aurilia:1978,DuffvanN:1980,AurNicTown:1980,HenTeit:1986,Aurilia:1991}.
Let
\vspace{-3mm}\be
F = \sdfrac{1}{\,4!}\, F\!_{\a\b\g\l}\, dx^\a \wedge dx^\b \wedge dx^\g \wedge dx^\l
\label{F4}
\vspace{-3mm}\ee
be a $4$-form field strength which is the curl of a totally anti-symmetric $3$-form gauge potential
\vspace{-4mm}\be
F\!_{\a\b\g\l} =4\, \pa_{[\a} A_{\b\g\l]} = 4\, \na_{[\a} A_{\b\g\l]} = \na_{\a} A_{\b\g\l} - \na_{\b} A_{\a\g\l}+  \na_{\g}A_{\a\b\l}- \na_{\l} A_{\a\b\g}
\label{divF}
\vspace{-2mm}\ee
so that
\vspace{-3mm}\be
F= dA\,,\qquad A= \sdfrac{1}{\,3! } \, A_{\a\b\g} \ dx^\a \wedge dx^\b \wedge dx^\g
\label{FdA}
\vspace{-2mm}\ee
i.e.~$F$ is an exact $4$-form. As a natural generalization of ordinary electromagnetism where $F = dA$ is an exact $2$-form exterior
derivative of the $1$-form vector gauge potential $A = A_\m \,dx^\m$, let $F$ be provided with the `Maxwell' action
\vspace{-2mm}\be
S\!_F =  -\sdfrac{1}{2\vk^4}\int F \wedge \st F=  -\sdfrac{1}{\,48\, \vk^4}  \int d^4x \sqrt{-g} \  F\!_{\a\b\g\l}F^{\a\b\g\l}
= \sdfrac{1}{2\vk^4} \int d^4x \sqrt{-g}\, \tF^{\,2}
\label{Maxw}
\vspace{-2mm}\ee
where
\vspace{-2mm}\be
\tF \equiv \st F = \sdfrac{1}{\,4!}\, \ve_{\a\b\g\l}\, F^{\a\b\g\l}\,, \qquad F\!_{\a\b\g\l} = -\ve_{\a\b\g\l}\,\tF
\label{Fdual}
\vspace{-2mm}\ee
is the scalar Hodge star $\st$ dual to $F$, {\it cf.}~(\ref{Hodgestar}), and $\vk$ is a free parameter whose significance
as the topological susceptibility of the gravitational vacuum is discussed in appendix~\ref{App:Susc}.

Now the point is that when the rank of the $D$-form $F$ is matched to the number of $D = 4$ spacetime dimensions, the free
`Maxwell' theory~(\ref{Maxw}) has two very special properties, namely: 
\vspace{-3mm}
\begin{enumerate}[label={(\roman*)},labelsep=5pt,itemsep=-2mm,leftmargin=30pt ]
\item $F$ is constrained to be a constant, with no propagating degrees of freedom, and 
\item its stress tensor $T^{\m\n}_{  F}$ is proportional to the metric $g^{\m\n}$, hence equivalent to a
cosmological term. 
\end{enumerate}
\vspace{-3mm}
The simplest example of this is usual $2$-form electrodynamics in $D = 2$ spacetime  dimensions, where
the classical Maxwell action is
\vspace{-3mm}\be
-\frac{1}{4e^2} \int d^2x \sqrt{-g}\, F\!_{\a\b}F^{\a\b}= \frac{1}{2e^2} \int d^2x \sqrt{-g}\, \tF^{\,2}
\label{Max2D}
\vspace{-1mm}\ee
with the dual $\tF =  \frac{1}{2} \ve_{\a\b}F^{\a\b} =  F^{01}$ the electric field in one spatial dimension. The stress tensor
corresponding  to~(\ref{Max2D}) is $-g^{\m\n} \tF^{\,2}/2e^2$,  provided $F\!_{\a\b}  = \pa_\a A_\b  -  \pa_\b A_\a$ is taken to be
independent of the metric, with spacetime indices raised by $g^{\a\b}$. Since the electric field $\tF$ is constrained by Maxwell's
eqs.~$\pa_\n F^{\m\n} = j^\m$ in $D = 2$, to be a spacetime constant of integration, i.e.~$\tF = const$.~in the
absence of sources $j^\m = 0$, the stress tensor of electric field energy, proportional to $g^{\m\n}$ is equivalent to a cosmological
vacuum energy. Classical Maxwell theory contains no propagating degrees of freedom at all in one space
plus one time dimension, in the absence of sources, and the constant $\tF$ simply parametrizes the energy of the vacuum.
In $D = 2$ the electric charge $e$ has mass dimension one, while in $D = 4$, the constant $\vk$ also carries mass
dimension one if the field strength tensor $F\!_{\a\b\g\l}$ is mass dimension four.

The exactly analogous situation obtains in $D = 4$ for~(\ref{F4})--(\ref{Fdual}), and in fact may be generalized to any even
$D$ spacetime dimension~\cite{HenTeit:1986}. The equivalence to $\La$ in $D = 4$ follows once again the fact that the
energy-momentum-stress tensor corresponding to~(\ref{Maxw})
\vspace{-3mm}\be
T^{\m\n}_{  F} = \frac{2\!}{\!\!\sqrt{-g}} \frac{\d S\!_F}{\d g_{\m\n}} =
-\frac{1}{4!\, \vk^4}\left(\sdfrac{1}{2} g^{\m\n} F^{\a\b\g\l}F\!_{\a\b\g\l} - 4 F^{\m\a\b\g}F^{\n}\!_{\a\b\g}\right)
= -\frac{1} {2\vk^4}\, g^{\m\n} \,\tF^{\,2}
\label{TF}
\vspace{-2mm}\ee
is proportional to the metric tensor, if the convention that $F\!_{\a\b\g\l}$ with all lower indices is independent of the metric is
again adopted. Analogous to the $D = 2$ case that $\tF$ is a constant follows from the sourcefree `Maxwell' eq.~obtained
by variation of~(\ref{Maxw}) with respect to $A_{\a\b\g}$, viz.
\vspace{-4mm}\be
\na\!_\l F_{ \a\b\g}\!^{\,\l}  =  0 \,,\qquad {\rm for} \qquad J^{\a\b\g}=0
\label{FnoJ}
\vspace{-4mm}\ee
and $\pa_\l \tF\!=\!0$, so that $\tF \!= \!\tF_0$ is a spacetime constant -- in the complete absence of any sources $J = 0$.

Hence~(\ref{F4})--(\ref{Fdual}), and (\ref{TF})--(\ref{FnoJ}) are completely equivalent to a cosmological term 
in Einstein's eqs.~in $D = 4$ dimensions, with the identification
\vspace{-3mm}\be
\La_{\rm eff} = \frac{4\p \GN}{\vk^4}\, \tF^{\,2} \ge 0
\label{Lameff}
\vspace{-3mm}\ee
the effective (necessarily non-negative) cosmological constant term for $\vk$ and $\tF =  \tF_0$ real constants.
In this way one can freely trade a positive cosmological constant $\La$ of classical GR for a new fundamental constant $\vk$ 
of the low energy EFT, together with an integration constant of the constraint $\pa_\l \tF= 0$.

It may seem at first sight that little has been gained by this trade of an equivalent reformulation of the $\La$ term as a $4$-form
gauge field at the classical level. However the free integration constant $\tF_0$ can then be fixed by a classical global boundary
condition in flat space, without any reference to quantum zero-point energy, UV divergences, or cutoffs. A vanishing $\La_{\rm eff}$
corresponds instead to the vanishing of the sourcefree `electric' field strength $\tF = F^{0123}$ in infinite three-dimensional
empty flat space, analogous to the vanishing  of $\tF =  F^{01}$ electric field of one space dimension in the absence of sources.
In either case this is simply the classical state of lowest energy, as well as the unique state that is even under the discrete
symmetry of space parity inversion.

Moreover this setting of the value of the free constant $\tF = \tF_0$, which is \emph{a priori} independent of geometry, to zero in empty
flat space is \emph{required} by the sourcefree Einstein's~eqs.
\vspace{-4mm}\be
\left[R_{\m\n} - \sdfrac{R}{2} g_{\m\n}\right]_{\rm flat} = 0 =-\La_{\rm eff}\Big\vert_{\rm flat}\,\h_{\m\n}
\label{Einflat}
\vspace{-2mm}\ee
viewed as a low energy EFT. This shows already that flat space QFT estimates of vacuum energy in any way
dependent upon UV cutoffs or heavy mass scales are \emph{inconsistent} with Einstein's eqs. It is well-known 
that QFT in flat space is sensitive only to \emph{differences} in energy. Hence the absolute value of quantum zero 
point energy in flat space, and its dependence upon cutoffs or UV regularization schemes is arbitrary and of no physical 
significance, a point made before, e.g.~\cite{Sola:2022}, but well worth emphasizing. The value of $\La_{\rm eff}$ is 
significant only through its gravitational effects, and hence cannot be evaluated in isolation, but only within the context of 
a gravitational EFT, `on shell' as in~(\ref{Einflat}), and only if each side of (\ref{Einflat}) can be evaluated \emph{independently}. 
This becomes possible only if $\La_{\mathrm{eff}}$ is a \emph{free} constant of integration, not a fixed parameter of the Lagrangian.

What has been gained then is that whereas in the usual treatment of $\La$ as a fixed parameter of the classical theory~(\ref{Scl}), 
which receives quantum corrections, dependent upon zero point energies and is apparently sensitive to UV physics, treating 
$\La_{\rm eff}$ instead as an integration constant of the classical `Maxwell' eq.~$\pa_\l \tF = 0$ of a $4$-form gauge field,
the free constant $\tF = \tF_0$ is independent of UV physics. Initially independent also of spacetime curvature, its value is  
uniquely determined by evaluating both sides of~(\ref{Einflat}) in flat space. If $\La_{\rm eff}$ is given by~(\ref{Lameff}), 
$\tF$ and $\La_{\rm eff}$ necessarily vanish in the flat space limit of GR, if classical (or semi-classical) GR is to be a consistent 
approximation to the low energy EFT of gravity.

Since the condition $\tF_0  =  0$ in flat space holds for any value of the EFT parameter $\vk$ in $D = 4$, which
remains arbitrary, the condition on the value of the integration constant $\tF_0$ at the minimum of energy does not involve
any fine tuning of fundamental constants or naturalness problem in the EFT with the $\La$ term replaced by~(\ref{Maxw}),
any more than it does for setting the electric field strength $F^{01}\!=\! 0$ in $D = 2$ classical Maxwell theory.
Parameters of the EFT Lagrangian, such as $\vk$ (or $\La$), may receive UV divergent contributions at
higher loop order that require UV regularization and renormalization, but the value of the classical `electric' field 
and integration constant $F^{0123}$ in flat space does not.

Thus the reformulation~(\ref{Lameff}) of the cosmological constant $\La_{\rm eff}$ in terms of $\tF =\tF_0$ and $\vk$ shifts the
consideration of cosmological vacuum energy away from the UV divergences of QFT to a macroscopic (IR) boundary condition
solving the classical constraint eq.~(\ref{FnoJ}) and minimization of energy in flat space. Although very simple
mathematically, and a completely equivalent parametrization of the $\La  \to  \La_{\rm eff}$ term in the classical Einstein
eqs.~in the absence of any sources for $F$, trading $\La$ for $\tF$ and a boundary condition through~(\ref{Lameff}) is
a significant step conceptually. For in addition to removing the fine tuning or naturalness problem of $\La$, introducing
an independent $4$-form field $F$ in place of constant $\La$ also allows for the introduction of sources in~(\ref{FnoJ}) that
will enable $F$ (and hence $\La_{\rm eff}$) to change, departing from its zero value in infinite sourcefree flat space
in finite calculable ways, and eventually to become a full-fledged dynamical variable of the low energy EFT in its own~right.

\section{The Chern-Simons $3$-Form and Anomaly Current Source for $F$}
\label{Sec:CS}

In section~\ref{Sec:F} the $4$-form field strength $F=dA$ was postulated as an independent degree of freedom, with 
the observation that it contributes to the EFT of low energy gravity in the same way as an effective vacuum energy and 
cosmological term according to~(\ref{Lameff}). This has the advantage of reformulating the naturalness problem of a
vanishing cosmological term as simply the solution of the sourcefree `Maxwell' eq.~(\ref{FnoJ}) that minimizes the 
`electric' energy in flat space, required by consistency of GR in its flat space limit~(\ref{Einflat}). In this section a 
fundamental geometric origin of $F$ is proposed by identifying the abelian $3$-form potential $A$ with the possible
torsion dependent part of the Chern-Simons $3$-form defined by the topological Euler-Gauss-Bonnet term in the the trace
anomaly~(\ref{tranom}). This identification determines the source current for the `Maxwell' eq.~(\ref{FnoJ}) in terms 
of the conformalon scalar $\vf$, that allows $F$ and hence the vacuum energy $\La_{\rm eff}$ to change.

Of the several terms in the trace anomaly, the Euler-Gauss-Bonnet invariant $E$ in~(\ref{ECdef}) is distinguished by its 
topological character. Its integral is a topological invariant insensitive to local variations, and therefore can be related to global
macroscopic effects, analogous to the index theorems associated with the $\ve_{\a\b\m\n} F^{\a\b}F^{\m\n}$ topological density 
of the axial anomaly~\cite{Bertlbook}. Just as the axial anomaly density can be expressed as the total divergence of a gauge 
dependent Chern-Simons current, the topological character of $E$ implies that it is also is a total divergence of a topological 
current, $E=\na\!_\m \W^\m$, with $\W^\m$ dependent upon the choice of local Lorentz frame, through the $S\!O(3,1)$ gauge 
connection.

The explicit form of the topological $3$-form gauge field and $\W^\m$ associated with $E$ follows from its relation to the $4$-form
field strength
\vspace{-4mm}\be
\mathsf{F}\equiv \e_{abcd}\, \mathsf{R}^{ab}\wedge  \mathsf{R}^{cd} = \sdfrac{1}{4} \e_{abcd}\, R^{ab}_{\ \ \,\a\b}\, R^{cd}_{\ \ \,\g\l}
\, dx^\a \wedge dx^\b \wedge dx^\g \wedge dx^\l
\label{Fdef}
\vspace{-5mm}\ee
by the Hodge star $\st$ dual operation
\vspace{-4mm}
\begin{align}
\st \mathsf{F}& =  \ve^{\a\b\g\l} \left(\sdfrac{1}{4} \e_{abcd}\, R^{ab}_{\ \ \,\a\b}\, R^{cd}_{\ \ \,\g\l}\right)
=  \sdfrac{1}{4} \e_{abcd}\,\e^{mnrs}\, R^{ab}_{\ \ \ mn}R^{cd}_{\ \ \ rs}   \nonumber \\
& =-\left(R_{abcd}R^{abcd} - 4 R_{ab}R^{ab}+ R^2\right)  =  -E
\label{starF}
\end{align}
\vspace{-1.4cm}

\noindent
where the Latin $a,b,\dots$ are tangent space indices and the Greek $\a,\b,\dots$ are spacetime coordinate indices respectively.
The overall minus sign in~(\ref{starF}) is the result of Lorentzian metric signature cf.~(\ref{esign}).

The curvature $2$-form $\mathsf{R}^{ab}$ in~(\ref{Fdef}) is defined by the Cartan structure eq.
\vspace{-4mm}\be
 \mathsf{R}^{ab} = d\w^{ab} +\w^{ac} \wedge \w^{\ \,b}_c \equiv \sdfrac{1}{2} R^{ab}\!_{\,\m\n}\, dx^\m\wedge dx^\n
\label{Car1}
\vspace{-3mm}\ee
that gives the Riemann curvature components $R^{ab}_{\ \ \,\m\n}$ in terms of the affine connection $1$-form
\vspace{-5mm}\be
\w^{ab} = -\w^{ba} = \w^{ab}\!_{\,\m}\, dx^\m
\vspace{-5mm}\ee
which specifies the law of parallel transport of orthonormal frames in tangent space. Thus $\w^{ab}$ may be regarded
as a local gauge potential for the Lie algebra of the $\cG =S\!O(3,1)$ Lorentz group, in close analogy to Yang-Mills gauge
potentials for any internal group $\cG$, for which $\mathsf{R}^{ab}$ would be the $2$-form field strength tensor.

The $4$-form $\mathsf{F}$ dual to $E$ by~(\ref{starF}) is exact, i.e.~$\mathsf{F}=\mathsf{dA}$, 
where $\mathsf{A}$ is the $S\!O(3,1)$ Lorentz frame dependent Chern-Simons $3$-form~\cite{PadYal:2011}
\vspace{-4mm}\be
\mathsf{A}= \e_{abcd}  \left(\w^{ab} \wedge d \w^{cd} + \sdfrac{2}{3}\, \w^{ab} \wedge \w^{ce}\wedge \w^{fd}\, \h_{ef}\right)
\label{eA}
\vspace{-4mm}\ee
which has the spacetime coordinate components
\vspace{-4mm}\be
\mathsf{A}_{\a\b\g} =  3!\, \e_{abcd}  \left(\w^{ab}\!_{\,[\a}\, \pa\!\!_{\ \b}\, \w^{cd}\!_{\,\g]}
+ \sdfrac{2}{3}\, \w^{ab}\!_{\,[\a}\,\w^{ce}\!_{\,\b}\, \w^{fd}\!_{\,\g]}\, \h_{ef}\right)
\label{Acomp}
\vspace{-4mm}\ee
completely anti-symmetrized in its three indices $\a,\b,\g$. These relations imply, cf.~(\ref{deps})
\vspace{-2mm}\be
E =  -  \big(\st d\mathsf{A}\big) = \sdfrac{1}{\,3! } \, \ve^{\a\b\g\m} \pa_\m \mathsf{A}_{\a\b\g}
=\sdfrac{1}{\,3! } \,\,\sdfrac{1}{   \sqrt{-g}}\,\pa_\m \left( \sqrt{-g}\, \ve^{\a\b\g\m} \mathsf{A}_{\a\b\g}\right)
= \sdfrac{1}{\,3! } \,\na\!_\m \left( \ve^{\a\b\g\m}\mathsf{A}_{\a\b\g}\right)
\label{EdA}
\vspace{-2mm}\ee
demonstrating that the integrand $\sqrt{-g}\, E$ is in fact a total derivative of a coordinate frame dependent abelian current
dual to $\mathsf{A}$, with $\W^\m  = \ve^{\a\b\g\m} \mathsf{A}_{\a\b\g}/3!$, analogous to the topological density 
$\ve_{\a\b\m\n} F^{\a\b}F^{\m\n}$ of the axial anomaly which is the total derivative of the gauge dependent Chern-Simons 
current. Eq.~(\ref{EdA}) holds in the absence of torsion, since the it has been so far implicitly assumed that the connection 
$\w^{ab}$ is the usual Riemannian or Levi-Civita connection of the metric $g_{\m\n}$.

To show that $\mathsf{A}$ is indeed an abelian gauge field, consider the response of the gauge connection to an infinitesimal local
$S\!O(3,1)$ tangent frame rotation
\vspace{-4mm}\be
\d_{\th}\, \w^{ab} = 2\, \w^{c[a}\,\th^{\,b]}\!_{\,c} + d\th^{\,ab} \,,\qquad \d_{\th}\, \mathsf{R}^{ab}
 = 2\, \mathsf{R}^{c[a}\,\th^{\,b]}\!_{\,c}
\label{gauge}
\vspace{-4mm}\ee
where $\th^{ab}(x) = - \th^{ba}(x)$ are the $6$ functions of the local Lorentz frame transformation. A short exercise then shows that
the gauge potential $3$-form $\mathsf{A}$ of~(\ref{eA}) transforms under~(\ref{gauge}) as the exact differential
\vspace{-4mm}\be
\d_{\th}\, \mathsf{A} = \e_{abcd}\,   d\th^{\,ab} \wedge d \w^{cd} = d \left( \e_{abcd}\,  \th^{\,ab} d \w^{cd}\right)
\label{delA}
\vspace{-4mm}\ee
and hence $\d_\th\,\mathsf{F} = 0$, so that the corresponding abelian field strength tensor is invariant under the particular form of
the local $2$-form gauge parameter $\Th  =   \e_{abcd}\,  \th^{ab}\, d \w^{cd}$. This gauge transformation establishes the
Chern-Simons $3$-form $\mathsf{A}$ as an abelian gauge potential. Since in order for the transformation~(\ref{delA}) to be 
non-null, $\Th$ itself must be co-exact, {\it i.e.}~not itself an exact $2$-form, the gauge transformation~(\ref{delA}) has just $3$ 
independent components. This means that of the $4$ independent components of $\mathsf{A}$, three of them are pure gauge, 
and only one is gauge invariant under~(\ref{gauge}). Hence the single scalar $\st \mathsf{F}$ carries the full gauge invariant 
field content of the $3$-form $\mathsf{A}$.

Now the essential point is that although $E$ is given in terms of the metric and its derivatives in Riemannian spacetime 
by (\ref{ECdef}), the $3$-form potential $\mathsf{A}$ given by (\ref{eA}) is defined in terms of the $S\!O(3,1)$ spin 
connection $\w^{ab}$ in an orthonormal basis, and {\it a priori independently} of the spacetime metric $g_{\m\n}$. 
This distinction becomes clear  when one considers Cartan's second equation of structure
\vspace{-4mm}\be
\mathsf{T}^a = de^a + \w^a\!_{\,b} \wedge e^b =  \sdfrac{1}{2}\, T^a\!_{\,bc}\, e^b\wedge e^c
= \sdfrac{1}{2}\, T^a\!_{\,\m\n}\, dx^\m\wedge dx^\n\
\label{Car2}
\vspace{-4mm}\ee
which defines the torsion $2$-form $\mathsf{T}^a$~\cite{Hehl:1976,EguGilHan:1980}.
This definition may be solved algebraically for the spin connection, {\it viz.}~\cite{Hehl:1976}
\vspace{-4mm}\be
\w_{ab\,\m} = -\w_{ba\,\m} = \y^\n\!_{\, a}\, \h_{bc}\, \big(\na_\m e^c\!_{\ \n}\big) -  K_{abc}\, e^c\!_{\ \m} 
\label{affcon}
\vspace{-2mm}\ee
whose first term is a purely Riemannian part in terms of the torsionless covariant derivative
\vspace{-5mm}\be
\na\!_\m e^c_{\ \n} = \pa_\m e^c\!_{\,\n} - \G^\l_{\ \, \m\n} e^c_{\ \l}
\label{natet}
\vspace{-5mm}\ee
with respect to the symmetric Levi-Civita connection $\G^\l_{\ \,\m\n}$ of~(\ref{Chris}), which depends upon the metric 
and its derivatives.  The second part of (\ref{affcon}) is dependent upon the torsion through the contorsion tensor
\vspace{-3mm}\be
K_{abc} = \sdfrac{1}{2}\,\big(T_{abc} + T_{bca} - T_{cab}\big)
\label{cont}
\vspace{-1mm}\ee
which is defined by (\ref{Car2}) independently of the metric. The definitions and properties of the vierbein field $e^a\!_{\,\m}$ 
and its inverse $\y^\m\!_{\,a}$ used here are given by eqs.~(\ref{genmet})--(\ref{prop}) of appendix~\ref{App:Tetrad}.

It follows from (\ref{affcon}) that if all components of torsion vanish,  $T^a\!_{\,bc} = 0$, then the affine connection
$\w_{ab\,\m}$ reduces to the Levi-Civita connection, which in holonomic coordinates is just the usual symmetric Riemann-Christoffel 
symbol $\G^\l\!_{\ \m\n}$ of~(\ref{Chris}), that is fully specified by the metric and its derivatives. In that case of vanishing torsion
the Chern-Simons $3$-form (\ref{eA}) is purely Riemannian $\mathsf{A} =\mathsf{A_R}$ and metric dependent in the standard way. 
Conversely, if torsion (\ref{Car2}) is non-vanishing, both $\w_{ab\,\m}$ and $\mathsf A$ will contain a torsion dependent 
part $\mathsf{A_T}$,  which may be treated as dynamical variables that can be varied independently of the metric $g_{\m\n}$. 
In the general case (\ref{eA}) will contain both purely Riemannian and torsion dependent terms, so that
\vspace{-5mm}\be
\mathsf{A}  = \mathsf{A_R}+ \mathsf{A_T}\,,\qquad E= -(\st d \mathsf{A_R})
\label{Adecomp}
\vspace{-5mm}\ee
since the the spin connection (\ref{affcon}) itself by which $\mathsf{A}$ is defined contains purely Riemannian and 
torsional terms. In (\ref{Adecomp}) $E$ is understood  to be the purely Riemannian, torsionless Euler-Gauss-Bonnet 
integrand of (\ref{ECdef}), derived from $\mathsf{A_R}$ alone, so that $E= -(\st d \mathsf{A_R})$ replaces (\ref{EdA}) 
when torsion is present and $\mathsf{A_T} \neq 0$.
\vspace{-2mm}

The independent variation of the affine connection is the basis of the first order or Palatini formalism of GR~\cite{MTW,Ferraris:1982}. 
In the case of the EH action this first order formalism leads to $T^a\!_{\,bc}=0$ in the absence of spin currents, and
hence turns out to be equivalent to the more common approach to GR where the connection is fixed to be the torsionless Christoffel connection~(\ref{Chris}) from the start. For more general actions, including that of the conformal  anomaly, independent variation of the 
affine connection and the metric generally leads to different Euler-Lagrange eqs., so that the resulting Einstein-Cartan theory differs 
in general from the torsionless theory.
\vspace{-2mm}

In the EFT based on the conformal anomaly it is only the particular dependence on the spin connection through
the $3$-form gauge field $\mathsf{A}$ of (\ref{eA}) that enters, which has only $4$ (not the $24$ of $\w_{ab\,\m}$) 
independent components. Hence rather than adopting the full first order formalism, the proposal for the EFT is that the 
torsional parts of the Chern-Simons $3$-form (\ref{eA}) and $4$-form (\ref{Fdef}) be identified with the corresponding quantities 
introduced in Sec.~\ref{Sec:F}, i.e.
\vspace{-5mm}\be
A = \mathsf{A_T}\qquad {\rm and} \qquad F = \mathsf{F_T}
\label{Aident}
\vspace{-5mm}\ee
with $E$ replaced by $E -(\st d\mathsf{A_T})= E -\tF$ in the anomaly effective action (\ref{Sanom}), and with
$A=\mathsf{A_T}$ treated as an independent variable, to be varied independently of the spacetime metric $g_{\m\n}$. 
From (\ref{Car2})-(\ref{cont}) this independent variation is possible since (\ref{eA}) holds also in a general Einstein-Cartan 
spacetime, if there is no {\it a priori} condition on the torsion, which is defined independently of the metric.
\vspace{-2mm}

With (\ref{EdA}), the decomposition (\ref{Adecomp}), and the identification~(\ref{Aident}), the term in the anomaly effective 
action linear in $E$ and the conformalon $\vf$ is replaced by $(E- \tF)\vf$, and the second term can be integrated by parts, so that
\vspace{-3mm}\be
\sdfrac{b'}{2\,} \int d^4 x \sqrt{-g}\, \big(E- \tF\big)\,\vf = \sdfrac{b'}{2\,} \int d^4 x \sqrt{-g}\, E\,\vf
-\sdfrac{b' }{2} \sdfrac{1}{\,3! } \,\int   d^4x \sqrt {-g}\,  A_{\a\b\g}\, \ve^{\a\b\g \m}\, \pa_\m \vf
\label{Ephi}
\vspace{-3mm}\ee
up to a surface term which does not affect local variations and may be taken to vanish for suitable boundary conditions.
Then defining the $3$-current
\vspace{-5mm}\be
J^{\a\b\g} \equiv  -\sdfrac{\,b' }{2}\, \ve^{\a\b\g \m}\, \pa_\m \vf
\label{Jphi}
\vspace{-5mm}\ee
the last term in~(\ref{Ephi}) can be expressed in the form
\vspace{-4mm}\be
S_{\rm int}[\vf, A] = \sdfrac{1}{\,3! }\int  d^4x \sqrt {-g}\ J^{\a\b\g} A_{\a\b\g}
\label{Sint}
\vspace{-4mm}\ee
analogous to a $J\cdot A$ interaction of ordinary electromagnetism.
\vspace{-2mm}

The analogy with electromagnetism is apt because the current~(\ref{Jphi}) is covariantly conserved, i.e.~
\vspace{-3mm}\be
\na\!_\g J^{\a\b\g} = -\sdfrac{\,b'}{2}\, \frac{1 }{  \sqrt{-g}}\ \pa_\g \left (\sqrt{-g}\, \ve^{\a\b\g\m}\, \pa_\m \vf\right ) =
-\sdfrac{\,b'}{2}\, \ve^{\a\b\g\m}\, \pa_\g\pa_\m \vf  = 0
\label{Jcons}
\vspace{-2mm}\ee
since $\sqrt{-g} \,\ve^{\a\b\g \m}$ is independent of the spacetime metric cf.~(\ref{deps}).
Thus the $J\cdot A$ interaction~(\ref{Sint}) is invariant under the abelian gauge transformation
\vspace{-4mm}\be
\d_\Th A = d \Th \,\qquad \d_\Th F = d^2\Th = 0
\vspace{-4mm}\ee
where $\Th$ is an arbitrary $2$-form, by another integration by parts. This means that~(\ref{Jphi}) is a candidate source term
for the `Maxwell' eq.~(\ref{FnoJ}), consistent with abelian gauge invariance, in analogy with ordinary electromagnetism.

With $A$ defined by the identification (\ref{Aident}) in terms of the spin connection {\it a priori} independent of the spacetime metric, 
it is a dynamical variable of the EFT in its own right. Its variation independently of $g_{\m\n}$ is just what is required to arrive at 
`Maxwell' eqs.~for $F$ with the conserved current (\ref{Jphi}) as their source. Thus identifications of the $3$-form $A$ and 
$4$-form $F = dA$ by (\ref{Aident}), with the interaction~(\ref{Sint}) deduced from the anomaly effective action have 
the consequence that the vacuum energy defined by $\La_{\rm eff}$ of~(\ref{Lameff}) will change if and when $\vf$ does 
and $J^{\a\b\g} \neq 0$, provided torsion (\ref{Car2}) and $\mathsf{A_T} \neq 0$.
 
\section{The Effective Theory of Gravity in the Absence of Torsion}
\label{Sec:EFTI}

Assembling the elements of the previous sections,  the effective action for low energy gravity in the absence of any
torsional contribution to the Chern-Simons $3$-form (\ref{eA}) is
\vspace{-4mm}
\begin{align}
S^{\rm (I)}_{ \rm eff}[g;\vf;A] = \sdfrac{1}{16\p \GN} \! \int\!   d^4 x \sqrt{-g}\, R + S\!_{\cA}[g;\vf]   + S\!_F[g;A] \hspace{1cm}\rm (I)
\label{Seff1}
\end{align}
\vspace{-1.2cm}

\noindent
where
\vspace{-6mm}
\begin{align}
 S\!_{\cA}[g; \vf] &=  \sdfrac{b' }{2\,} \int  d^4 x\sqrt{-g}\,\bigg\{\! - \left(\sq \vf\right)^2
+ 2\,\Big(R^{\m\n}  -  \tfrac{1}{3} Rg^{\m\n}\Big) \,(\na_\m\vf)\,(\na_\n\vf)\bigg\}   \nonumber \\
&\quad + \sdfrac{1}{2}  \int  d^4 x\sqrt{-g} \left\{ b\, C^2 +b' \left(E - \tfrac{2}{3}\sq R\right)+ \sumi \,\b_i\, \cL_i\right\}\vf 
\end{align}
\vspace{-1cm}

\noindent
is the conformal anomaly effective action (\ref{Sanom}) and $S\!_F$ is the `Maxwell' action (\ref{Maxw}) of the $4$-form gauge field $F$,
with~(\ref{divF}). In this case (I) the $3$-form gauge field $A$ and $4$-form field strength $F = dA$ are not coupled
in any way to the anomaly effective action $S\!_\cA[g;\vf] $. Hence $\tF = \tF_0$ is sourcefree and constant, as in~(\ref{FnoJ}),
and entirely equivalent to an effective cosmological term $\La_{\rm eff}$ by~(\ref{Lameff}).

With the condition that $\tF_0 = 0$ in asymptotically flat space, the lowest energy ground state, $\tF$ and $S\!_F[A;g]$
then drop out entirely.  In this case the EFT (I) is just classical GR with $\La_{\rm eff} \!=\! 0$ and with the addition of the
conformal anomaly effective action $S\!_\cA[g;\vf]$. Alternately, $S\!_F[A;g]$ with non-zero constant $\tF$ may be
retained to parametrize an arbitrary positive constant vacuum energy $\La_{\rm eff}$ in some region(s) of space, its value
to be determined by appropriate boundary conditions, as in the application to gravitational condensate stars of section~\ref{Sec:GravCond}.

The classical Euler-Lagrange eqs.~following from variation of~(\ref{Seff1}) are
\vspace{-3mm}\be
\D_4 \vf \equiv  \na_\m \left(\na^\m\na^\n +2R^{\m\n} - \tfrac{2}{3} R g^{\m\n} \right) \na_{ \n}\vf
= \sdfrac{1}{2}\left(E - \sdfrac{2}{3}\sq R\right) + \sdfrac{1}{2b'}\, \left( b C^2 +  \sumi \,\b_i\cL_i\right) \hspace{3mm}{\rm (I)}
\label{phieom1}
\vspace{-4mm}\ee
for the conformalon scalar $\vf$, together with  the semi-classical Einstein eq.
\vspace{-3mm}\be
\hspace{1cm}R_{\m\n} - \sdfrac{1}{2} R g_{\m\n} = 8\p \GN \left( T\!_{F\,\m\n} +T\!_{\cA\,\m\n}[g;\vf] + T^{cl}_{\m\n}\right)
\label{Eineq}
\vspace{-3mm}\ee
with $T^{\m\n}_{  F}$ and $\La_{\rm eff}$ given by~(\ref{TF}) and~(\ref{Lameff}), and with
\vspace{-3mm}\be
T_{ \cA}^{\,\m\n}\,[g;\vf] \equiv \sdfrac{2 }{   \sqrt{-g}} \, \sdfrac{\d}{\d g_{\m\n}} \ S\!_\cA[g;\vf]
= b' E^{\m\n} + bC^{\m\n} + \sumi\ \b_i T^{(i)\,\m\n}
\label{Tphi}
\vspace{-3mm}\ee
the stress tensor resulting from the metric variation of $S\!_\cA[g;\vf]$. The first contribution here is
\vspace{-5mm}
\begin{align}
E_{\m\n}&=- 2\,(\na_{(\m}\vf) (\na_{\n)} \sq \vf)  + 2\na^\a \big[(\na_\a \vf)(\na_\m\na_\n\vf)\big]
- \tfrac{2}{3}\, \na_\m\na_\n\big[(\na_\a \vf)(\na^\a\vf)\big]\nonumber \\
& +\tfrac{2}{3}\,R_{\m\n}\, (\na_\a \vf)(\na^\a \vf) - 4\, R^\a_{\ (\m}\left[(\na_{\n)} \vf) (\na_\a \vf)\right]
 + \tfrac{2}{3}\,R \,(\na_{(\m} \vf) (\na_{\n)} \vf) \nonumber \\\notag
&\hspace{-5mm}+ \tfrac{1}{6}\, g_{\m\n}\, \left\{-3\, (\sq\vf)^2 + \sq \big[(\na_\a\vf)(\na^\a\vf)\big]
+ 2\, \big( 3R^{\a\b} - R g^{\a\b} \big) (\na_\a \vf)(\na_\b \vf)\right\} \\
& \hspace{-1.8cm}   - \tfrac{2}{3}\, \na_\m\na_\n \sq \vf  - 4\, C_{\m\ \n}^{\ \,\a\ \b}\, \na_\a \na_\b \vf
- 4\, R^\a\!_{ (\m} \na_{\n)} \na_\a\vf  + \tfrac{8}{3}\, R_{\m\n}\, \sq \vf   +\tfrac {4}{3}\, R\, \na_\m\na_\n\vf
- \tfrac{2}{3} \left(\na\!_{(\m}R\right) \na_{\n)}\vf \nonumber \\
&\hspace{-1mm}+ \tfrac{1}{3}\, g_{\m\n}\, \left[ 2\, \sq^2 \vf + 6\,R^{\a\b} \,\na_\a\na_\b\vf
- 4\, R\, \sq \vf  + (\na^\a R)\na_\a\vf\right]
\label{Eab}
\end{align}
\vspace{-1.3cm}

\noindent
which is the metric variation of all the $b'$ terms in~(\ref{Sanom}), both quadratic and linear in
$\vf$~\cite{EMVau:2006,EMZak:2010,EMSGW:2017}, while
\vspace{-7mm}
\begin{subequations}
\begin{align}
& C_{\m\n}  \equiv  -\frac{1}{\sqrt{-g}\ } \frac {\d }{\d g^{\m\n}} \left\{\int d^4x\sqrt{-g}\,C^2\,\vf\right\}
=-4\,\na_\a\na_\b\,\Big( C_{(\m\ \n)}^{\ \ \a\ \ \b} \,\vf \Big)  -2\, C_{\mu\ \,\nu}^{\ \,\a\ \,\b}\, R_{\a\b}\, \vf\\
  &\hspace{3.8cm} T^{(i)}_{\m\n} \equiv  -\frac{1}{\sqrt{-g}\ }\, \frac{\d}{\d g^{\m\n}}\left\{ \int d^4x\sqrt{-g}\,  \cL_i\, \vf\right\}
\end{align}
\label{CFterms}\end{subequations}
\vspace{-1cm}

\noindent
are the metric variations of the last two $b$ and $\b_i$ terms in~(\ref{Sanom}), both of which are only linear in $\vf$.

In~(\ref{Eineq}) a classical matter/radiation stress tensor $T^{cl}_{\m\n}$ independent of $\vf$ has been allowed as well.
If any $\b_i\cL_i$ or $T^{cl}_{\m\n}$ terms are non-zero, the Euler-Lagrange eqs.~of the fields (or fluid constitutive relations)
upon which these additional degrees of freedom depend must be appended to~(\ref{phieom1}) and~(\ref{Eineq}),
to close the system~(\ref{phieom1})--(\ref{CFterms}).

Note that the stress tensor $T\!_\cA^{\ \m\n}[g;\vf]$ derived from $S\!_\cA[g;\vf]$ with $\vf$ treated
as a classical field, is the finite renormalized stress tensor of the underlying quantum conformal QFT, after all short distance
UV divergences have been removed, with $\big\lag T^{\m\n}\big\rag_{\mathrm{flat}}$ in infinite flat space defined to be zero, by
the consistency condition~(\ref{Einflat}), where $\vf$ may be taken to vanish as well. If boundary conditions different
from infinite empty flat space are considered, $T^{\m\n}_F$ and $\La_{\rm eff}$ of~(\ref{TF}) parametrize any
finite vacuum energy in place of $\La$ by~(\ref{TF}), and $\vf$ may be different from zero. For example the stress
tensor~(\ref{Eab}) with $\vf \propto z^2/a^2$,  $\sq^2 \vf  = 0$ and $\La_{\rm eff} = 0$ can
account for the Casimir energy and force between two parallel plates a distance $a$ apart in the $z$ direction
in flat space, when boundary conditions appropriate to that situation are imposed on $\vf$~\cite{EMVau:2006}.

The scalar-tensor theory (\ref{Seff1}), based on first principles of QFT and the conformal anomaly SM fields, is quite distinct from other 
modifications of GR, such as Brans-Dicke theory, Hordenski theory or massive gravity. As the possible anomaly sources for $\vf$
in~(\ref{phieom1}) are  negligibly small in our local neighborhood, this effective theory easily passes the most constraining 
solar system tests, as well as the laboratory bounds on \emph{ad hoc} modified gravity or other scalar-tensor
theories~\cite{Will:2014,EMSGW:2017}. Being generally covariant, the theory described by~(\ref{Seff1})--(\ref{CFterms}) is 
consistent with the Weak Equivalence Principle (WEP) and local Lorentz invariance. The solutions of~(\ref{phieom1}) propagate at 
$c$, the speed of both light and of gravitational waves, consistent with present gravitational wave observations.

There are nevertheless two situations where the effects of~(\ref{Tphi}) are significant at macroscopic scales and lead
to effects qualitatively different from the purely classical theory:
\begin{enumerate}[label=(\roman*), itemsep=-2mm]
\vspace{-2mm}
  \item In the vicinity of horizons, where there are large local blueshifts and the light cone singularities of anomalies come to the fore, relevant
    for BHs and cosmology, {\it cf.}~section~\ref{Sec:GravCond};
\item When the source for $\vf$ in~(\ref{phieom1}) is non-gravitational in origin and sufficiently strong, such as from the QCD trace anomaly
$\cL_G$ in dense nuclear matter, where scalar gravitational waves may be generated in compact binary mergers
and in the hot, dense early universe~\cite{EMSGW:2017}.
\vspace{-2mm}
\end{enumerate}

Since the eq.~of motion~(\ref{phieom1}) for $\vf$ involves the fourth order conformal Panietz-Riegert operator $\D_4$
of~(\ref{Deldef})~\cite{Panietz,Rieg:1984}, and typically, differential eqs.~higher than second order possess negative energy
and/or unstable solutions growing in time, a few additional comments are in order here. Note first that $S\!_\cA$ and $\D_4$ do not occur
in isolation, but as part of the EFT of gravity, subject to the first class constraints of diffeomorphism invariance. These constraints restrict
the class of physically allowed solutions both classically and quantum mechanically. Since $S\!_\cA$ is quadratic in $\vf$, a simple case
in which the specific effects of the higher derivative terms in the anomaly effective action can be studied is the exactly solvable limit of
$\GN^{-1}  \to  0$, where the EH term is neglected, and on the product space ${\mathbb R} \times {\mathbb S}^3$, where
$E, C^2$ and $\sq R$ all vanish. The result of this analysis is that only a small subset of solutions survive the constraints of
diffeomorphism invariance, and correspondingly, only a small subspace of physical states with positive norm, and no propagating modes
whatsoever, survive quantization~\cite{StatesAntMazEM:1997}.

When $\GN^{-1}  \neq  0$ and the EH term of classical GR is added, the resulting EFT~(\ref{Seff1}) becomes non-linear, but
then can be studied in linearized perturbation theory around flat space. The mixing of the scalar $\vf$ with the conformal factor 
of the metric turns this constrained mode of classical GR into a propagating one. The result is that~(\ref{Seff1})--(\ref{Eab}) 
predicts the existence of scalar gravitational waves of positive energy~\cite{EMSGW:2017}, i.e.~a `breather' mode polarization 
in addition to the two transverse, traceless gravitational wave modes of classical GR. The active linearized solutions of~(\ref{phieom1}) 
are those with $\sq^2 \vf \!=\! 0$, but $\Y \!=\! \sq \vf \neq  0$. In other words, the modes with $\sq \vf \!=\! 0$ decouple entirely, 
and the remaining solutions satisfy the second order eq.~$\sq \Y \!=\!0$, with positive energy. Thus in linearization around flat space,  
half of the solutionsof the fourth order $\D_4$ operator are eliminated by the constraints, and do not appear in the physical asymptotic 
states and $S$-matrix of the EFT. There is no instability in this case either, consistent with general results on the stability of flat space 
to quantum corrections~\cite{MazEM:NPB1990,AndMolEM:2003}.

Since $S\!_\cA$ is derived from the conformal anomaly of well-behaved SM fields, each with a unitary $S$-matrix in flat space, 
the EFT incorporating the anomaly is not expected to lead to unphysical instabilities in weakly curved or asymptotically flat space 
at low energies, within its range of validity. While this interesting issue certainly deserves further investigation, and should be
revisited now with the introduction of the $4$-form gauge field, the detailed studies of the EFT obtained from the conformal 
anomaly to date are consistent with this physical expectation.

\section{The Effective Theory of Gravity in the Presence of Torsion}
\label{Sec:EFTII}

In the presence of torsion the Chern-Simons $3$-form potential (\ref{eA}) acquires a torsional dependent term $\mathsf{A_T}$,
which may be identified with the $3$-form potential of Sec.~\ref{Sec:F} and makes the additional contribution (\ref{Sint}) to
the effective action, which in this case becomes
\vspace{-3mm}
\be
S^{\rm (II)}_{ \rm eff}[g;\vf;A] = \sdfrac{1}{16\p \GN} \! \int\!   d^4 x \sqrt{-g}\, R + S\!_\cA[g;\vf]  + S\!_F[g;A]
+ S\!_{\rm int}[\vf; A]  \hspace{1cm} \rm (II)
\label{Seff2}
\vspace{-3mm}\ee
in which $S\!_{\rm int}$ is given by~(\ref{Sint}). Since $A=\mathsf{A_T}$ is independent of the metric, the independent variables
of~(\ref{Seff2}) are $(g_{\m\n}, \vf, A_{\a\b\g})$. 
\vspace{-1mm}

A possible additional torsion dependent term arising from the conformal anomaly of $N_f$ massless fermions 
is~\cite{BuchOdinShap,Shap:2002,CamShap:2022} 
\vspace{-6mm}
\be
S\!_W [g;\vf;W] = - \sdfrac{N_f}{48\p^2}  \int  d^4 x\sqrt{-g}\, \big(\na_\m W^\perp_\n- \na_\n W^\perp_\m\big)^{\! 2} \vf
\label{SW}
\vspace{-5mm}
\ee
in terms of the transverse part of the axial vector field
\vspace{-5mm}\be
W\!_\m = \tfrac{1}{4} \,\e_{abcd}\,K^{abc}\,e^d_{\ \m} = \tfrac{1}{8}\, \e_{abcd}\,T^{abc}\,e^d_{\ \m}
\label{Wdef}
\vspace{-4mm}\ee
dependent upon torsion. This term has been omitted from (\ref{Seff2}), assuming that it can be varied independently of 
$(g_{\m\n}, \vf, A_{\a\b\g})$, and the resulting Euler-Lagrange eq.
\vspace{-4mm}\be
\na\!_\n\,\big(\na^\n W^\perp_\m \vf\big) =0
\label{Wperp}
\vspace{1mm}\ee
admits the solution $W^\perp_\m=  0$. 
\vspace{-3mm}

Since $\sqrt{-g}\, \ve^{\a\b\g \m}$ appearing in~(\ref{Sint}) is in fact independent of the metric, {\it cf.}~(\ref{deps}),
the additional term $S_{\rm int}$ makes no contribution to the stress tensor, and variation of~(\ref{Seff2}) with respect to the metric 
gives the Einstein eqs., identical in form to~(\ref{Eineq}).
\vspace{-2mm}

Variation of (\ref{Seff2}) with respect to $\vf$ yields
\vspace{-3mm}
\be
\D_4 \vf  = - \sdfrac{1}{2}\, \tF + \sdfrac{1}{2}\left(E - \sdfrac{2}{3}\sq R\right)+ \sdfrac{1}{2b'}\, \left(b C^2 +  \sumi \,\b_i\cL_i\right)
 \hspace{1.2cm}{\rm (II)}\hspace {-2cm}
\vspace{-2mm}\label{phieom2}
\ee
instead of~(\ref{phieom1}). Since the `Maxwell' action~(\ref{Maxw}) and interaction term~(\ref{Sint}) may be varied 
independently with respect to $A_{\a\b\g}$, with $g_{\m\n}$ and $\vf$ held fixed, the novel feature of (\ref{Seff2}) is the `Maxwell' eq.
\vspace{-4mm}\be
\na_{\!\l} F^{\a\b\g\l} = \vk^4  J^{\a\b\g} = -\sdfrac{\vk^4 b'}{2}\,\ve^{\a\b\g\l}\,\pa_\l\vf
\label{Maxeq}
\vspace{-2mm}\ee
with the source current~(\ref{Jphi}).  Upon taking its dual, with~(\ref{Fdual}), this becomes
\vspace{-2mm}\be
\pa_\l\left(\tF - \sdfrac{\vk^4 b'}{2}\, \vf \right) = 0
\label{paF}
\vspace{-3mm}\ee
which is an eq.~of constraint  that is immediately solved by
\vspace{-4mm}\be
\tF = \sdfrac{ \vk^4 b' }{2} \,\vf + \tF_0
\label{Fsoln}
\vspace{-3mm}\ee
in which $\tF_0$ is a spacetime constant. Thus $\tF$ can be eliminated in favor of $\vf$ and (\ref{phieom2}) becomes
\vspace{-2mm}\be
\D_4 \vf + \sdfrac{ \vk^4 b' }{4} \,\vf =-\sdfrac{\tF_0}{2}  + \sdfrac{1}{2}\left(E - \sdfrac{2}{3}\sq R\right)
+ \sdfrac{1}{2b'}\, \left( b C^2 +  \sumi \,\b_i\cL_i\right) \,.\hspace{5mm} \rm (II) \hspace{-1cm}
\label{phieom3}
\vspace{-2mm}\ee
For asymptotically flat boundary conditions $\tF_0$ may be set to zero, but is retained here for more general cases.
\vspace{-2mm}

For the second form of the effective action (\ref{Seff2}), the result is that $\tF$ given 
by (\ref{Fsoln}) is no longer a constant, and will change, as will the effective cosmological term 
$\La_{\rm eff}$, when $\vf$ changes according to (\ref{phieom3}).  Eq.~(\ref{phieom3}) for $\vf$ and the
Einstein eq.~(\ref{Eineq}) together with  (\ref{TF}) and (\ref{Tphi})-(\ref{CFterms}) is the form of the 
proposed EFT of low energy gravity and dynamical vacuum energy in the presence of torsion.
\vspace{-2mm}

In deciding which which form of the EFT (\ref{Seff1}) or (\ref{Seff2}) applies, the critical question is whether spacetime 
acquires a non-vanishing torsion. This question remains open at present. Since the vanishing of torsion is equivalent 
to the vanishing of the covariant derivative of thevierbein, according to the definition (\ref{Car2}), it has been suggested 
that a natural place for torsion to appear is where the vierbein vanishes, and the locking together of the $S\!O(3,1)$ tangent 
space gauge group and $GL(4,\mathbb{R})$ group of coordinates transformations is broken~\cite{DauriaRegge:1982}. 
This hypothesis will be adopted in the application of the following section.

\section{Gravitational Vacuum Condensate Stars in the EFT of Gravity}
\label{Sec:GravCond}

In~\cite{gravastar:2001,MazEMPNAS:2004,MazEM:2015} it was proposed that the solution of the multiple BH paradoxes is that the
final state of complete gravitational collapse is a gravitational condensate star rather than a BH.  The proposed gravastar is a compact
object with a physical surface of positive surface tension replacing the BH horizon, and a static region of de Sitter space with the eq.~of
state $p \!=\! -\r$ replacing the singular interior of a BH. Because such an object is both horizonless and non-singular, with low entropy,
it suffers from no information paradox, and is consistent with quantum unitary evolution. The EFT of Sec.~\ref{Sec:EFTII} provides a 
fundamental first principles Lagrangian basis for this proposal.

The requirements for a gravitational condensate star to be realized in gravitational collapse are first, that quantum vacuum polarization effects
can grow large in the vicinity of a BH horizon and second, they can induce a phase transition to a non-vanishing interior gravitational Bose-Einstein
condensate (GBEC) with $p \!=\! -\r$, equivalent to a non-zero $\La_{\rm eff}$. 

That the anomaly effective action $S\!_\cA[g;\vf]$ and
stress tensor~(\ref{Eab})--(\ref{CFterms}) of section~\ref{Sec:EFTI} can have substantial effects in spacetimes with horizons, satisfying
this first part of the gravastar hypothesis may be seen in the case of the exterior static Schwarzschild spacetime
\vspace{-4mm}\be
ds^2 = -f(r)\, dt^2 +  \sdfrac{dr^2}{h(r)}+  r^2\left( d\th^2 + \sin^2 \th\,d\f^2\right)
\label{sphsta}
\vspace{-3mm}\ee
with
\vspace{-2mm}\be
f(r) = h(r) = 1-\sdfrac{\rM}{r} = 1 - \sdfrac{2\GN M}{c^2\,r}\,.
\label{Sch}
\vspace{-1mm}\ee
The general solution to~(\ref{phieom1}) for $\vf \!=\! \vf(r),\, \tF \!=\!\tF_0 \!=\!0$ and $\cL_i\!=\!0$ that is finite as $r\! \to\!  \infty$ 
for the Schwarzschild metric~(\ref{sphsta})--(\ref{Sch}) was found previously to be~\cite{EMVau:2006,EMZak:2010}
\vspace{-3mm}
\begin{subequations}
\begin{align}
\sdfrac{d \vf_{_S}}{\!dr}   & =\sdfrac{c_{_S} \rM}{r(r-\rM)} - \sdfrac{2}{3\rM\!\!} \left(\sdfrac{r}{\rM\!\!} + 1 + \sdfrac{\rM\!}{r} \right)
\ln   \left(1-\sdfrac{\rM\!}{r}\right)   - \sdfrac{\!2}{3\rM\!\!} - \sdfrac{1}{r} \\[1ex]
\vf_{_S}(r) & = c_{_S}\ln \left(1-\sdfrac{\rM\!}{r}\right) 
+ \int_{r/\rM}^\infty   dx \, \Bigg\{\sdfrac{2}{3x}\left(x^2+ x+ 1\right)
 \ln \left(1-\sdfrac{1}{x}\right) + \sdfrac{2}{3} + \sdfrac{1}{x}\Bigg\}
\label{phiSint}
\end{align}
\label{phiS}\end{subequations}
\vspace{-1.1cm}

\noindent
in terms of the dimensionless integration constant $c_{_S}$.
\!\!\footnote{An additional constant of integration $c_\infty$ in $\vf_{\!S}$ has been dropped here and in (\ref{philim}), since it 
does not contribute to the anomaly stress tensor (\ref{Tphi}).} This solution has the limits
\vspace{-3mm}\be
\vf_{_S}(r) \to \left\{ \begin{array} {lr} c_{_S}\ln   \left(1-\sdfrac{\rM\!}{r}\right) + c_1
- 2  \left(1-\sdfrac{\rM\!}{r}\right) \left[\ln  \left(1-\sdfrac{\rM\!}{r}\right) - \sdfrac{1}{6}\right] + \dots \,,\quad & r \to \rM\\[2ex]
- \left(c_{_S} + \sdfrac{11}{9} \right) \sdfrac{\rM\!}{r} - \left(2 c_{_S} + \sdfrac{13}{9}\right) \sdfrac{\rM^2}{4r^2} + \dots\,, & r \to \infty
\end{array}\right.
\label{philim}
\vspace{-2mm}\ee
where the constant $c_1$ is the finite integral in~(\ref{phiSint}) evaluated at the lower limit $r = \rM, x = 1$. Substituting the solution~(\ref{phiS}) into the anomaly stress tensor~(\ref{Tphi})--(\ref{Eab}), one finds
\vspace{-3mm}\be
\big(T^\m_{ \ \,\n}\big)_{\!\cA} \to \frac{c_{_S}^2}{6\rM^2\!}\ \frac{b'}{(r-\rM)^2}\ {\rm diag}\,(-3,1,1,1) \to \infty
\qquad {\rm as} \qquad r \to \rM
\label{ThorizM}
\vspace{-2mm}\ee
diverging on the Schwarzshild BH horizon for any $c_{_S} \!\neq\! 0$.

This leading order $(r-\rM)^{-2}$ divergence of the stress tensor on the horizon may be understood from the kinematic
blueshifting of local frequencies in~(\ref{Sch}) near the horizon according to
\vspace{-4mm}\be
\w_{\rm loc}(r) =\,\sdfrac{\w_\infty\!}{\!\!\!\sqrt{f(r)}}
\label{locw}
\vspace{-2mm}\ee
relative to that at $r\! = \!\infty$. The corresponding energy $\hbar \w_{\rm loc}(r)$ diverges as $r  \to  \rM$ and therefore becomes
much greater than any finite mass scale. This is reflected in the fact that the wave eq.~for a quantum field of arbitrary finite mass
and spin becomes indistinguishable from that of a massless conformal field in the horizon limit $r  \to  \rM$~\cite{EMZak:2010},
and hence the conformal anomaly effects come to the fore. Since the stress tensor is a dimension four, conformal weight four operator,
it behaves generically as the fourth power of $\w_{\rm loc}$ in~(\ref{ThorizM}), {\it i.e.}~$\propto f^{-2}$. Noting that $\vf$ is a scalar,
as is the norm of the static Killing field $K = \pa_t$ of~(\ref{sphsta}), which is $\sqrt{-K^\m K_\m} = \sqrt{-g_{tt}} = \sqrt{f(r)}$,
the divergence of~(\ref{ThorizM}) depending on the inverse fourth power of this norm is also a coordinate invariant scalar, and observer
independent.

The diverging behavior of the local stress tensor~(\ref{ThorizM}) shows that the anomaly stress tensor can become important near the
horizon of a BH and even dominate the classical terms in the Einstein eq.~(\ref{Eineq}), the smallness of the curvature
tensor there notwithstanding. Even with $c_{_S} \!=\! 0$ in~(\ref{phiS}), which can be arranged by specific choice of the state
of the underlying QFT, to remove the leading $f^{-2}$ divergence in~(\ref{ThorizM}), there remain subleading divergences
proportional to $f^{-1}, (\ln f)^2$ and $\ln f$. In fact, there is no solution of~(\ref{phieom1}) in Schwarzschild spacetime with
$\vf = \vf(r)$ only, corresponding to a fully Killing time $t$ invariant and spherically symmetric quantum state, with a finite stress
tensor at both regular singular points $r = \rM$ and $r=\infty$ of the differential eq.~(\ref{phieom1}). This result, following simply
and directly from the conformal anomaly effective action, confirms results of previous studies of the stress tensor expectation value in
specific states in Schwarzschild spacetime~\cite{ChrFul:1977}.

The divergences on either the future or past BH horizon (but not both) can be cancelled by allowing linear time dependent
solutions of~(\ref{phieom1}), which give rise to a Hawking flux $\lag T^t_{\ \,r}\rag$, such as in the Unruh states; or by relaxing the regularity
condition at infinity which gives rise to a non-zero stress tensor there,  as in the Hartle-Hawking state. This thermal state is
both incompatible with asymptotically flat boundary conditions and unstable. Usually the \emph{assumption} of regularity of
the semi-classical $\lag T^\m_{\ \,\n}\rag$ on the horizon is used to argue for the necessity of Hawking radiation flux
$\lag T^t_{\ \,r}\rag > 0$~\cite{FredHag:1990}. However the converse is also true: if a \emph{truly static} and stable asymptotically
flat solution of the final state of gravitational collapse is sought with $\lag T^t_{\ \,r}\rag = 0$, then quantum effects at the horizon,
specifically due to the conformal anomaly $T\!_\cA^{\ \m\n}$ cannot be neglected, and imply instead the breakdown of regularity there.

A similar behavior is observed in de Sitter spacetime, and indeed in any spherically symmetric static spacetime~(\ref{sphsta}) 
with a horizon at which $-K^\m K_\m  =  f(r)  \to  0$. With $\La$ positive, the static patch of de Sitter space is of
this form with $f(r)  \propto  h(r)  =  1 - H^2r^2 =  1  -  \La r^2/3$. The general spherically symmetric static solution of~(\ref{phieom1})
for $\vf  =  \vf(r)$ which is regular at the origin in this case is~\cite{EMVau:2006,EMZak:2010}
\vspace{-3mm}
\begin{align}
\vf_{dS}(r) & =  \ln\left(1-H^2r^2\right) + c_0 + \frac{q}{2}
\ln\left(\frac{1-Hr}{1+Hr}\right) + \frac{2c_{_H} - 2 - q}{2Hr} \ln\left(\frac{1-Hr}{1+Hr}\right)\nonumber \\
& \to  \Big[c_{_H} + \Big(c_{_H} -1 - \sdfrac{q}{2}\Big)  \big(1-Hr\big) + \dots \ \Big ]\, \ln\big(1-Hr\big) + c' + \cO\, (1-Hr)
\label{phidS}
\end{align}
\vspace{-1.2cm}

\noindent
where the constant $c'=  c_0 +  (2-c_{_H})\ln 2$. Substituting this into the anomaly stress tensor~(\ref{Tphi}) gives
\vspace{-3mm}\be
\big(T^\m_{ \ \,\n}\big)\!_\cA \to \sdfrac{2}{3}\, c_{_H}^2\, H^4 \frac{b'}{(1-Hr)^2}\ {\rm diag}\,(-3,1,1,1) \to \infty
\qquad {\rm as} \qquad r \to \rH \equiv H^{-1}
\label{ThorizH}
\vspace{-2mm}\ee
which also diverges as $f^{-2}$ for any $c_{_H}  \neq  0$ as the de Sitter static horizon is approached.

As in the Schwarzschild case this divergence can be removed if a thermal state is considered, but then only if the temperature is
precisely matched to the Hawking temperature $T_{_H}\!=\! H/2\p$ associated with the horizon, which in the de Sitter case
leads to the maximally $O(4,1)$ symmetric state~\cite{CherTag:1968,BunDav:1978}. However, this state is not a 
vacuum state of QFT and is unstable to particle pair creation, much as a uniform, constant electric field
is~\cite{EM:1985,Polyakov:2008,AndEM:2014a,AndEM:2014b,AndEM:2018}, and for essentially the same reason. 
Due to the non-existence of a global static time by which positive and negative frequency (particle and anti-particle) solutions can 
be invariantly distinguished, a time independent Hamiltonian bounded from below and stable vacuum state cannot be defined. The horizon
where $f\!=\! 0$ and the Killing vector of time translation $\pa_t$ becomes null in either Schwarzschild or de Sitter space is the sign of
this, so that $\La$ cannot be globally constant and positive everywhere in space in QFT. The conformal anomaly shows this through its
sensitivity to lightlike correlations on the horizon and non-local boundary conditions on the quantum state. Only flat space
with $\La_{\rm eff} \!=\! 0$ can be a candidate stable ground state of the semi-classical EFT, as the state of lowest energy with
vanishing `electric' field $\tF=0$, a global static Killing time and no horizon, as also required by the low energy EFT consistency
condition~(\ref{Einflat}).

With $\La_{\rm eff}$ replaced by a dynamical condensate according to~(\ref{Lameff}), vacuum energy can be non-zero only if localized
in space, within the static patch $r  <  \rH = H^{-1}$ of de Sitter space. If one seeks a stable spherically symmetric static solution of the
EFT, the light cone enhanced effects of the anomaly stress tensor~(\ref{anomT}) at both the Schwarzschild and de Sitter
horizons,~(\ref{ThorizM}) and~(\ref{ThorizH}) respectively, should be taken into account, and the assumption that either 
horizon is a mathematical boundary only should be re-examined. In the gravastar proposal~\cite{gravastar:2001,MazEMPNAS:2004}, 
the BH horizon is the location of a physical surface phase boundary layer between two different phases characterized by different 
values of the vacuum energy $\La_{\rm eff}$, regarded as a gravitational condensate. On general thermodynamic grounds the 
Gibbs relation $\r + p  \!=\!  sT + \m n \!=\!  0$ implies that the eq.~of state of a zero temperature condensate with no conserved particle 
number should be $p \!=\! -\r$. An argument based on non-relativistic condensed matter analogs given in~\cite{ChaplineHCL:2001} 
reached a similar conclusion.

In~\cite{MazEM:2015} the gravitational condensate star was shown to follow directly from Schwarzschild's
constant density interior solution in the limit $r\! \to\! \rM$, with $f(r) = h(r)/4$ leading to equal and opposite surface gravities
of the surface at $\rM \!=\! \rH$.  The surface tension of this physical boundary layer replacing the Schwarzschild and de Sitter horizons
is determined from the $\d$-function discontinuity in gradients of the surface gravities there, and the First Law becomes a purely
mechanical relation of a gravastar with this surface tension, and the relativistic analog of the Rayleigh surface tension of a fluid droplet.

The singular behavior of both $\vf$ and the anomaly stress tensor at the Schwarzschild and de Sitter horizons
(\ref{philim}), (\ref{ThorizM}) and  (\ref{phidS}), (\ref{ThorizH}) is clearly associated with the vanishing of
the norm of the static Killing vector field $\pa_t$ in each case. The coordinate singularities at these static horizons 
coincide with the vanishing or divergence of the vierbein $e^0_{\ t}$ or $e^1_{\ r}$ respectively. This is
exactly the locus of a possible breakdown of the locking together of the $S\!O(3,1)$ tangent space gauge group 
and $GL(4,\mathbb{R})$ group of coordinates transformations, where torsion may be expected to arise~\cite{DauriaRegge:1982}.
Thus it is natural to describe this region where $f(r), h(r)\!\to\! 0$ by the EFT (II) of section~\ref{Sec:EFTII}.
This EFT (II) then provides a mechanism and Lagrangian description for the second part of the gravitational condensate
star proposal of~\cite{gravastar:2001,MazEMPNAS:2004,MazEM:2015}, by allowing the value of $\La_{\rm eff}$ to change
from exterior to interior of the gravastar, through (\ref{Lameff}), (\ref{Fsoln}) and (\ref{phieom3}).

The scalar $\tF$, dual to the $4$-form field strength of sections~\ref{Sec:F}--\ref{Sec:CS}, is a classical coherent field 
that provides an explicit realization of a gravitational condensate $\La_{\rm eff}$ interior. When coupled to the conformalon 
scalar through the $3$-form abelian current $J^{\a\b\g}$, concentrated on a three-dimensional extended world tube of 
topology ${\mathbb R} \times {\mathbb S^2}$ where $\pa_\m\vf$ grows large, $\vf, \tF$ and the condensate $\La_{\rm eff}$ 
all change rapidly in the radial direction. This is precisely the appropriate description of a thin shell phase boundary
layer of a gravastar with $\mathbb S^2$ spatial topology sweeping out a tube in spacetime.

One may now search for static, rotationally invariant solutions of the EFT eqs.~of Secs.~\ref{Sec:EFTI}-\ref{Sec:EFTII} which are 
asymptotically Schwarzshild-like with $\La_{\rm eff}$ vanishing in the exterior region, changing rapidly but continuously
near the Schwarzschild  $\rM = 2\GN M/c^2$ or de Sitter $\rH = H^{-1}$ classical horizons by (\ref{Fsoln}) and (\ref{phieom3}) 
in the phase boundary region, and then remaining nearly constant $3H^2$ in the interior region.  The blueshifting of local frequencies 
$\sim  f^{-\frac{1}{2}}$ as in~(\ref{locw}) leads to the $\vf$ field having an increasingly large radial derivative in the vicinity
of $\rM\simeq \rH$, so that (\ref{Jphi}) and (\ref{Tphi}) and the torsional effects become significant there.
The physical thickness of the phase boundary surface layer where the anomaly stress tensor (\ref{ThorizM}) or (\ref{ThorizH}) 
becomes significant and large enough to compete with the classical terms and where the EFT (II) of \ref{Sec:EFTII}
must be used is of the order of $\sqrt{\rM L_{Pl}}$. The effects of  this surface layer and regular de Sitter interior on binary 
BH mergers, gravitational waves, ringdown and `echoes' can then be investigated in the EFT.

\section{Discussion and Outlook: Vacuum Energy as a Dynamical Condensate}
\label{Sec:DDE}

In this paper an EFT of gravity has been proposed taking account of the most significant macroscopic light cone effects of
the conformal anomaly of massless or light SM fields. In this EFT $\La$ is no longer a fundamental constant, whose value 
appears to be sensitive to ultra high energy physics, but rather a dynamical condensate described by a classical $4$-form 
field strength and $3$-form abelian gauge potential of sections~\ref{Sec:F}--\ref{Sec:CS}.

The EFT based on the conformal anomaly~(\ref{tranom}) introduces two relevant scalar degrees of freedom to low energy gravity
beyond classical GR. This follows from the local form of the effective action of the anomaly $S\!_\cA[g; \vf]$ of~(\ref{Sanom}) in
terms of the scalar $\vf$ conformalon field, which satisfies a fourth order eq.~of motion~(\ref{phieom1})-(\ref{phieom2}).
Before the addition of $S\!_\cA[g; \vf]$, the conformal factor of the metric in GR is constrained to be non-propagating by 
the classical diffeomorphism constraints of Einstein's eqs. Once $S\!_\cA[g; \vf]$ is added to the effective action, the conformal part 
of the metric mixes with one of the two $\vf$ modes, and gives rise to scalar gravitational waves~\cite{EMSGW:2017}. 

Likewise before the identification (\ref{Aident}) of the potential $A$ of the $4$-form gauge field~(\ref{F4}) with the torsional part 
of the Chern-Simon $3$-form, $F$ is constrained to be a constant and simply equivalent to a cosmological constant term 
in Einstein's eqs.~by (\ref{Lameff}). Once the interaction $S\!_{\rm int}[\vf,A]$ is added to the effective action, the
$3$-form current $J$ of~(\ref{Jphi}) provides a source for $F$ through (\ref{Maxeq}), and $F$, hence  $\La_{\rm eff}$
becomes dynamical through $\vf$. It is this second scalar degree of freedom of $\vf$ in $S\!_\cA[g; \vf]$ that allows (and requires) 
the vacuum energy to become a dynamical variable through~(\ref{Lameff}) and~(\ref{Fsoln}).

Thus in the EFT  each of the two scalar conformalon degrees of freedom present in the anomaly action for $\vf$ mix with, and source
the previously classically constrained conformal factor of $g_{\m\n}$ and $F$, turning each into full-fledged dynamical
degrees of freedom.  Assuming the SM, the only new free parameter of the EFT is the constant $\vk$, a kind of gravitational
vacuum susceptibility cf.~appendix~\ref{App:Susc}, determining the coupling of the current $J$ to the gauge field $A$,
field strength $F$, and $\La_{\rm eff}$ of~(\ref{Lameff}).

For the longstanding problem of the cosmological term when QFT is coupled to gravity, in either form of the EFT the classical
state of minimum energy is that of vanishing $4$-form condensate: $\tF\! =\! \tF_0\! =\! 0$. By the identification of $\La_{\rm eff}$
in~(\ref{Lameff}), this condition automatically sets the value of the cosmological term to zero in infinite flat Minkowski space.
By simply allowing a flat Minkowski solution, this removes one oft-stated obstacle and `no-go theorem' to solution of the `cosmological
constant problem'~\cite{WeinbergRMP:1989,KlinkVol:2010}. Moreover $\La_{\rm eff} \!= \! 0$ is required by consistency with
the classical or semi-classical Einstein eqs.~(\ref{Einflat}) or~(\ref{Eineq}) in infinite flat Minkowski space, in the complete
absence of matter and radiation. The condition~(\ref{Einflat}) is independent of UV physics or cutoffs, as simply the unique classical
state of minimum energy with zero curvature and zero torsion, and the stable ground state of the EFT of low energy gravity.

Empty flat space with zero external fields is also the unique classical state where the total anomaly $\cA = 0$,
so that the WZ anomaly action~(\ref{Sanom})--(\ref{SanomWZ}) has an additional global shift symmetry under $\vf \to \vf  +  \vf_0$
in that state, and $F^{0123} = 0$ is also required by spatial parity invariance of the ground state. These enhanced global
or discrete symmetries may be regarded as replacing the 't Hooft naturalness criterion~\cite{tHooft:1979} for $\La_{\rm eff}=0$.

The consistency condition of $\La_{\rm eff} = 0$ on the classical condensate in the low energy EFT of gravity in flat space
indicates that estimates of vacuum energy as sensitive to UV physics are not applicable if GR, with or without the conformal anomaly
addition, is to be a well-behaved EFT at macroscopic scales. Arguments or estimates of the cosmological term in flat space as
proportional to the fourth power of a UV cutoff (which break Lorentz invariance) or the fourth power of all masses $\sumi m_i^4 \ln m_i$
in QFT (as in dimensional regularization) have no physical meaning in the absence of gravitation, as well as being in conflict
with observations even on non-cosmological, solar system scales~\cite{Martin:2012}.

Instead a way around the `naturalness' problem of the $\La$ term is to extend the EFT of gravity beyond classical GR to contain 
the additional scalar conformalon degree(s) of freedom inherent in the conformal anomaly effective action~(\ref{Sanom}), and 
to replace the fixed parameter $\La$ of the classical theory by $\La_{\rm eff}$ of~(\ref{Lameff}) in terms of a $4$-form gauge 
field $F$. As a result of the `Maxwell' eq.~$F$ satisfies, this abelian gauge field becomes a fully dynamical degree of freedom 
of low energy gravity by~(\ref{Maxeq}) and (\ref{Fsoln}). The setting of an integration constant to zero at the minimum of energy 
in flat space is a solution of the naturalness problem of the cosmological term that involves no fine tuning of any fixed parameters 
of the EFT Lagrangian~(\ref{Seff1}) of macroscopic~gravity.

At lowest order all fields $(g_{\m\n}, \vf, A)$ in~(\ref{Seff1})-(\ref{Seff2}) are treated as classical. Quantum loop corrections are
then to be computed by the usual EFT method of appending local terms as needed to absorb UV divergences in an expansion in powers
of $1/M_{pl}^2$, maintaining the physical meaning of the constants of the lowest order EFT. Thus at one-loop order the energy
of the vacuum will continue to be defined to be identically zero in flat space by the consistency condition~(\ref{Einflat}) on the
constant $\tF_0 = 0$, hence $\La_{\rm eff} = 0$ in the absence of any sources, and with $\vf = 0$. All formal divergences in
$\lag \hat T^{\m\n}\rag_{\mathrm{flat}}$ in purely flat space, quartic or otherwise, are treated as without physical significance
and removed by this consistency condition on $\tF_0$. There is no sensitivity of this free integration constant, or $\La_{\rm eff}$
defined in terms of it by~(\ref{Lameff}), on UV divergences or UV mass scales.

Following the usual logic of EFT, divergences of quantum loops require the introduction of additional local terms which are higher order in
powers of the Riemann curvature tensor and its derivatives, divided by higher powers of a UV scale, presumably the Planck
mass scale $M_{\mathrm{Pl}}$~\cite{Don:1994,Bur:2004,Don:2012}, which ultimately limits the range of applicability
of the low energy EFT. These higher order effects and their renormalization should be defined so as not to disturb the meaning 
of the low energy parameters at lowest order, such as $\GN$, or ground state boundary condition on the condensate, $\La_{\rm eff}\!=\! 0$ 
in flat space. 

Logarithmic divergences in curved space (regulated by any covariant method)  require the introduction of counterterms 
proportional to the local $R^2$ and $C^2$ curvature invariants, together with the finite logarithmic running of their dimensionless 
couplings. These terms and local terms involving still higher numbers of derivatives are not treated as fundamental, but rather as 
suppressed at energy scales far below the Planck energy $M_{\mathrm{Pl}}c^2$, remaining negligibly small at macroscopic distance 
scales much greater than $L_{\mathrm{Pl}}$, as consistent with existing EFT results~\cite{Don:1994,Bur:2004,Don:2012}.

One may also turn the EFT logic around, to conjecture that the important role of the conformal anomaly and anomalies in general
as windows into the UV, and exceptions to the usual decoupling hypothesis of EFT, may indicate that in the fundamental theory all
masses vanish and conformal invariance is restored -- broken perhaps only spontaneously, at asymptotically
high energies. Speculations of this kind for resolution of the naturalness problem of the cosmological term and possible
relation with that of the Higgs mass hierarchy have been advanced by a number of authors,
e.g.~\cite{Bard:1995,BJ:2003,MeisNic:2008,tHooft:2015,ShapShim:2019}. Although no clearly successful complete theory
has emerged from these speculations, the idea of fundamental conformal invariance and relation between these large hierarchy
problems and the properties of the quantum vacuum rather than UV physics remains intriguing.

That both $\La$ and Higgs hierarchies may be resolvable only by the consistent inclusion of gravity receives some support from
the EFT approach to vacuum energy and the cosmological `constant' as a dynamical condensate proposed in this paper. 
To the extent that $\pa_\l \tF \neq 0$, this dynamical condensate necessarily requires non-vanishing spacetime torsion. The role
of relaxing the torsionless condition of classical GR in describing the condensate by~(\ref{Aident}) as relevant to resolving 
BH singularities was anticipated in~\cite{DauriaRegge:1982}, where specific models generating torsion were proposed.
The possible extensions of the EFT proposed in this paper to generate torsion dynamically and self-consistently remain to
be explored. The coupling of fermions to a condensate with torsion through the spin connection and possible relation to 
neutrino mass generation is another intriguing direction for future research. The microscopic constituents of the gravitational 
Bose-Einstein condensate (GBEC) described by $\tF$ from which its superfluid nature is emergent remain to be elucidated~\cite{Volov,Mazur:2007}.

As a more immediate matter,  the description of $\La_{\rm eff}$ as a dynamical condensate of a $4$-form gauge field
in the EFT of sections~\ref{Sec:EFTI}-\ref{Sec:EFTII} makes possible calculations of vacuum energy in numerous applications, first and foremost
for a gravitational condensate star interior and surface. For cosmology, the universe as the interior of a gravastar realizes the hypothesis
made in section~\ref{Sec:Intro} of automatically relating the effective value of the vacuum energy $\La_{\rm eff}$ to $3H^2$, and
hence to the horizon Hubble scale $H^{-1}$, with no fine tuning. More realistic cosmological models,  with $\W_\La  < 1$, require
dynamical EFT solutions including matter and radiation, rather than a purely static de Sitter vacuum condensate.

The tying of the value of $\La_{\rm eff}$ to the Hubble scale is clearly relevant to the `cosmic coincidence problem' of the
$\La$CDM model, and immediately suggests a rather different set of possibilities for cosmological models, in which
spatial inhomogeneities and/or boundary conditions at the Hubble scale $H^{-1}$ play an important role. The EFT of the
conformal anomaly coupled to the $3$-form potential and $4$-form abelian field strength term presented in sections~\ref{Sec:EFTI}
and \ref{Sec:EFTII} thus provides a distinctly new framework for dynamical dark energy in cosmology based on fundamental theory.

From the form of the anomaly $\cA$ in~(\ref{tranom}), any deviation from exact homegeneity and isotropy will lead
in general to $F_{\m\n}F^{\m\n}  \neq  0$ for the photon radiation field, and ${\rm tr}\, \{G_{\m\n}G^{\m\n}\}  \neq  0$
for the electroweak and QCD color gauge fields in the unconfined phase of the early universe. This will induce changes in
the conformalon field $\vf$ through its eq.~of motion~(\ref{phieom1}), which will then cause the field strength $\tF$ and
hence the vacuum energy $\La_{\rm eff}$ to change. After the transition to the confining phase of QCD, baryonic matter
will still contain non-vanishing gluonic condensates and thus still act as a source for $\vf$, thereby coupling
non-relativistic baryonic matter to dynamical vacuum energy as well.

Thus although $\vf$ is \emph{not} an inflaton, it is a dynamical scalar that is well-grounded in QFT of the SM and can produce backreaction
effects on the vacuum energy when fluctuations away from exact homogeneity and isotropy are admitted. It permits interaction between
both radiation and matter with dynamical dark energy, in which adiabaticity of the matter and radiation components will no longer
be satisfied in general, in effect introducing a bulk viscosity into the cosmological fluid.  If $\La_{\mathrm{eff}}  \propto  \tF^2$ does 
not remain constant in the de Sitter phase, deviations from the $\La$CDM cosmological model are to be expected, and evolution away
from a pure de Sitter phase due to cosmological horizon modes~\cite{AndEM:2009} becomes calculable, and testable by
the cosmological data of large scale structure.

In addition to removing the singularity and paradoxes of BHs, developing detailed predictions from the EFT proposed
will allow study of gravastar stability, normal modes of oscillation, surface modes and `echoes' that can be tested
with gravitational wave and multi-messenger signals from binary merger events, in the increasing data samples expected
in the future. The prediction of scalar gravitational waves can also be tested by the coming global array of gravitational
wave antennae~\cite{HagEra_SGW:2020}.

The effects of rotational angular momentum have been neglected in the simplest gravastar solution, although the first
steps in including those effects have been taken in~\cite{BelGonEM1:2022,BelGonEM2:2022}. The EFT solutions of static or
stationary gravastars also leaves unexamined the process of their formation, and in particular the behavior of the stress
tensor near the would-be horizon of collapsing matter, which would have to activate the dynamical condensate terms
of~(\ref{Seff2})--(\ref{phieom2}). These and many other interesting questions remain to be addressed in the context of the
EFT of dynamical vacuum energy proposed in this paper.

\vspace{4mm}
\centerline{\bf Acknowledgements}
\vspace{2mm}

The author expresses his appreciation to his colleague Prof. Pawel O.~Mazur for a critical reading of this paper,
and for bringing reference~\cite{DauriaRegge:1982} to the author's attention, to Prof. Ilya L.~Shapiro for
useful discussions and bringing references~\cite{Shap:2002, Obuk:1983,BuchOdinShap:1985,CamShap:2022}
to his attention, and to Joan Sol\`a Peracaula for bringing reference~\cite{Sola:2022} to his attention as well.

\vspace{8mm}

\centerline{\bf References}
\vspace{-1.7cm}
\bibliographystyle{apsrev4-1}
\bibliography{gravity21Aug}

\appendix
\section{Metrics, Tetrads, Differential Forms, and Hodge Star Dual}
\label{App:Tetrad}

In this first Appendix conventions and mathematical details used in the text are collected and catalogued.
The metric and curvature conventions used in this paper are those of Misner, Thorne \& Wheeler~\cite{MTW}.
Greek indices are four-dimensional coordinate (holonomic) indices, while Latin indices refer to local othonormal tangent space.

In the tetrad or vierbein formalism the metric line element is written
\vspace{-4mm}\be
ds^2 = g_{\m\n}(x)\, dx^\m dx^\n = e^a \h_{ab} e^b
\label{genmet}
\vspace{-4mm}\ee
with $\h_{ab} =$ diag $(-1,1,1,1)$ the flat spacetime tangent space Minkowski metric tensor, and $e^a(x)$ are the $1$-forms
\vspace{-6mm}
\begin{align}
e^a = e^a_{\ \m}\, dx^\m \qquad {\rm satisfying} \qquad e^a_{\ \m}\, e^b_{\ \n}\, \h_{ab} = g_{\m\n}\,,\quad
g^{\m\n} \,e^a_{\ \m }\,e^b_{\ \n} = \h^{ab}\,.
\label{tetrad}
\end{align}
\vspace{-1.2cm}

\noindent
The dual basis of vectors $\by_a$ satisfy
\vspace{-8mm}
\begin{subequations}
\begin{align}
 e^a_{\ \m}  \y^\m_{\ b} & = \d^a_{ \ b}\\
 e^a_{\ \n}  \y^\m_{\ a} & = \d^\m_{ \ \n}\\
 \y^\m_{\ a}  \y^\n_{\ b}\, g_{\m\n} & =\h_{ab}\\
 \y^\m_{\ a}  \y^\n_{\ b} \h^{ab} & = g^{\m\n}
\end{align}
\label{prop}\end{subequations}
\vspace{-1.5cm}

\noindent
defining the orthonormal basis in tangent space. Coordinate indices are lowered (resp.~raised) by the metric tensor $g_{\m\n}$
(resp.~its inverse $g^{\m\n}$). Tangent space indices are lowered or raised by the flat Minkowski tensor $\h_{ab}$ or $\h^{ab}$.
The covariant derivative of the vierbein field in~(\ref{natet}) is defined with respect to the torsionless Levi-Civita connection,
that in holonomic coordinates is the familiar Christoffel symbol~(\ref{Chris}), which is specified entirely by the metric tensor
and its first derivatives.
\vspace{-1mm}

The exterior derivative operator $d$ maps the general $p$-form
\vspace{-3mm}\be
Q^{(p)} = \sdfrac{1}{p!}\, Q^{(p)}\!_{\m_1\dots\m_p}\, dx^{\m_1} \wedge \dots \wedge dx^{\m_p}
= Q^{(p)}\!_{[\m_1\dots\m_p]}\, dx^{\m_1} \wedge \dots \wedge dx^{\m_p}
\label{Psin}
\vspace{-4mm}\ee
into the $p+1$-form
\vspace{-3mm}\be
d Q^{(p)} = \frac{1}{p!}\, \frac{\pa Q^{(p)}\!_{\m_1\dots\m_p}}{\pa x^\l}\, dx^\l\wedge dx^{\m_1} \wedge \dots \wedge dx^{\m_p}
= \pa\!_{\ [\l}Q^{(p)}\!_{\m_1\dots\m_p]}\ {\pa x^\l}\wedge dx^{\m_1} \wedge \dots \wedge dx^{\m_p}
\vspace{-3mm}\ee
where the square brackets denote anti-symmetrization with respect to the enclosed indices, and both $p$ and $p+1$ must be
$\le D$ in $D$ dimensions, by the anti-symmetry of the wedge~product.

The Hodge $\st$ dual operator maps the $p$-form~(\ref{Psin}) into the $(4-p)$-form
\vspace{-4mm}
\be
\st Q^{(p)} =  \sdfrac{1}{(4-p)!} \,\sdfrac{1}{p!} \, \ve_{\m_1\dots\m_4}\, g^{\m_1\n_1}\dots g^{\m_p\n_p}\,
Q^{(p)}\!_{\n_1\dots\n_p}\, dx^{\m_{p+1}} \wedge \dots \wedge dx^{\m_4}
\label{Hodgestar}
\vspace{-3mm}
\ee
in $D = 4$ dimensions. We make use of the notation
\vspace{-3mm}\be
 \e_{abcd} =\left\{ \begin{array}{rc} +1 &\, {\rm if}\  (a,b,c,d) = P\!_{\mathrm{even}}\,(0,1,2,3)\\
-1 &  {\rm if}\  (a,b,c,d) = P\!_{\mathrm{odd}}\,(0,1,2,3)\\
0 & {\rm any\ two\ indices\ equal} \end{array} \right.
\label{LCt}
\vspace{-3mm}\ee
for the totally anti-symmetric Levi-Civita tensor in the tangent basis, where $P\!_{\mathrm{even}}$ and $P\!_{\mathrm{odd}}$ denote
even or odd permutation respectively of the four indices which are its argument. The corresponding tensor in the
coordinate basis is denoted by
\vspace{-5mm}\be
\ve_{\a\b\g\l} \equiv  \e_{abcd}\,e^{a}_{\ \,\a}\, e^{b}_{\ \,\b}\, e^{c}_{\ \,\g}\, e^{d}_{\ \,\l}
\label{epstensor}
\vspace{-4mm}
\ee
which is used to define the volume $4$-form
\vspace{-3mm}
\be
\st \mathds{1} =  \sdfrac{1}{\,4!} \,\e_{abcd}\, e^{a} \wedge e^{a} \wedge e^{c} \wedge e^{d} =
\frac{1}{\,4!}\,\ve_{\a\b\g\l}  \ dx^\a \wedge dx^\b\wedge dx^\g \wedge dx^\l
\label{Volform}
\vspace{-3mm}
\ee
dual to the unit scalar, and the $4$-volume element
\vspace{-3mm}
\begin{align}
\int   \st \mathds{1} &=  \int \ve_{txyz} \ dt\, dx\,dy\,dz=
  \int \e_{abcd}\,e^{a}_{\ \,t}\, e^{b}_{\ \,x}\, e^{c}_{\ \,y}\, e^{d}_{\ \,z} \ dt\, dx\,dy\,dz
 \nonumber \\
& =  \int  {\rm det}\,\big(e^a_{\ \m}\big) \,d^4x =  \int   \sqrt{-g} \,d^4x
\end{align}
\vspace{-1.2cm}

\noindent
where $x^\m = (t, x, y, z)$ are general spacetime coordinate labels (not necessarily Minkowski).

Since by~(\ref{LCt}) $\e_{0123}= 1$, raising all indices by use of the Minkowski metric $\h^{ab}$ changes its sign, so that
$\e^{0123} = -1$, which leads to
\vspace{-5mm}\be
\e^{abcd}\e_{mnrs} = - 4! \, \d^a\!_{\,[m}\, \d^b\!_{\,n}\, \d^c\!_{\,r}\, \d^{d\,}\!_{s]}
\label{esign}
\vspace{-3mm}\ee
and~(\ref{Fdual}) of the text, as well as
\vspace{-2mm}\be
\st \st Q^{(p)} = (-)^{p + 1} Q^{(p)}
\vspace{-2mm}\ee
for the double dual of a $p$-form. Since $\ve_{\a\b\g\l} \propto \sqrt{-g}$ it follows that $\ve^{\a\b\g\l} \propto 1/ \sqrt{-g}$ and
\vspace{-4mm}\be
\pa_\m \,\left( \ve^{\a\b\g\l} \sqrt{-g}\,\right) =0
\label{deps}
\vspace{-3mm}\ee
which also can be verified directly from $\ve^{\a\b\g\l}  =  g^{\a\a'}g^{\b\b'}g^{\g\g'}g^{\l\l'}\ve_{\a'\b'\g'\l'}$
and properties~(\ref{genmet})--(\ref{prop}), with the definition~(\ref{epstensor}). These properties of the $\ve$ tensor are used
at several points in the main text, e.g.~in~(\ref{Jcons}) and~(\ref{paF}) to convert covariant derivatives to coordinate partial 
derivatives and \emph{vice versa}. That $S\!_{\rm int}$ of (\ref{Sint}) with (\ref{Jphi}) is independent of the spacetime metric 
$g_{\m\n}(x)$, and hence makes no contribution to the Einstein eq.~(\ref{Eineq}), also follows from (\ref{deps}).

\section{Topological and Torsional Susceptibility of the Gravitational Vacuum}
\label{App:Susc}

In this appendix we consider the physical interpretation of the parameter $\vk$ in~(\ref{Maxw}),  as a kind of
torsional topological susceptibility of the gravitational vacuum. This depends upon the identification of the $3$-form 
gauge field and associated field strength $F$, and follows by close analogy with the cases of the topological susceptibility 
of the QED$_2$ vacuum, and chiral susceptibility of QCD in $D\!=\!4$.

In QED$_2$ $\lag \tF (x) \tF(y)\rag = e^2 \d^2(x-y)$, and its Fourier transform at $k^2\!=\! 0$ (or indeed any $k$ in the absence
of charged sources) is simply the constant $e^2$~\cite{Seiler:2002}. Likewise for the free action~(\ref{Maxw}) in $D\!=\!4$
flat space we have
\vspace{-6mm}
\begin{subequations}
\begin{align}
&\hspace{2cm}\lag \tF (x) \tF(y)\rag_0 = \vk^4\, \d^4(x-y)\qquad \qquad {\rm and}\\[1ex]
&\ch\!_{F,\,0}(k^2)  = \int d^4\!x\, e^{ik\cdot (x-y)} \,  \big\lag \tF(x) \tF(y)\big \rag_0 = \vk^4 = \ch\!_{F,\,0}(0)
\label{chiF}
\end{align}
\end{subequations}
\vspace{-1.2cm}

\noindent
a finite constant. Since this free correlator is computed with the source current of~(\ref{Jphi}) set to zero, it corresponds
to $b' \!=\!0$, where all the matter fields contribution to the conformal anomaly are neglected, as in the quenched
limit of QCD, where an analogous expression for the chiral susceptibility holds~\cite{Witten:1979,Venez:1979,VendiVecc:1980,Seiler:2002}.
Once matter vacuum polarization effects are taken into account $\ch\!_F(k^2) \!\neq\! \ch\!_{F,\,0}(k^2) $ will no longer be independent
of $k^2$. However the limit
\vspace{-5mm}\be
\lim_{k^2 \to \infty} \ch\!_F (k^2) = \lim_{k^2 \to \infty} \ch\!_{F,\,0}(k^2)= \vk^4
\label{chilim}
\vspace{-4mm}\ee
remains to reflect the local $\d^4(x-y)$ short distance correlator of the free action~(\ref{Maxw}).
\vspace{-2mm}

If one were to define the topological susceptibility of the Riemannian $E$
\vspace{-4mm}\be
\ch\!_E (k^2) = \int d^4\!x\, e^{ik\cdot x} \, \big\lag E(x) E(0) \big\rag
\label{chiE}
\vspace{-4mm}\ee
directly in terms of the curvature invariants one would encounter the correlator of two dimension-four operators, with the expected
short distance singularity of $1/x^8$ as $x\!\to\! 0$. Thus the integral in~(\ref{chiE}) is undefined in perturbative quantum gravity and
badly (in fact, quartically) UV divergent. To define it requires promoting the correlator to a distribution with $a_2\sq^2\d^4(x),\, a_1\sq \d^4(x)$ and  $a_0\,\d^4(x)$ local contact terms added with arbitrary finite coefficients $a_2,a_1, a_0$, as in~\cite{Seiler:2002}. 
This corresponds to adding three local counterterms to the action in order to make the three subtractions necessary to remove the
quartic, quadratic and logarithmic divergences from~(\ref{chiE}), and thence to obtain a finite renormalized result in terms of these three
finite but unknown parameters. In the quenched approximation where the remaining finite terms vanish, only the local $\d$-functions and
derivatives thereof remain, and for $k^2 = 0$, \, $\ch\!_{E,\,0}(0) \!=\! a_0 \!=\! \vk^4$. 
\vspace{-2mm}

When $\tF$ is identified with the torsional part of the topological density of the Euler class, as in (\ref{Aident}), $\vk^4$ in the action~(\ref{Maxw}) parametrizes the logarithmic short distance $\vk^4\,\d^4(x)$ renormalized singularity of 
this torsional density. Because of~(\ref{chiF})--(\ref{chilim}), this is the leading order effect of quantum gravitational vacuum
fluctuations at short distances that is physically relevant to the $k^2\! \to\!  0$ low energy (light cone) correlations of the EFT.
\vspace{-2mm}

Since $F$ involves just one derivative of the gauge potential $A$ in the low energy EFT, $F/\vk^2$ is a quantum
operator of mass dimension $2$ in terms of $A/\vk^2$, in contrast to $E$ which is fourth order in metric derivatives. By this accounting~(\ref{Maxw}) is a dimension $4$ (rather than dimension $8$) operator which is marginally IR relevant in the Wilsonian EFT 
sense in $D = 4$, just as~(\ref{Max2D}) is in $D = 2$. In the QED$_2$ Schwinger model case there are no UV divergences whatsoever
and $e^2$ is a UV finite coupling, despite having dimensions of $(mass)^2$. This may provide an interesting prototype of how parameters
with positive mass dimensions can nevertheless remain finite and insensitive to UV corrections. If the matter contributions to the
vacuum polarization self-energy $\int \! d^4\!x\, e^{ik\cdot(x-y)} \lag J^{\a\b\g}(x) J^{\l\m\n}(y) \rag$ are also UV finite, as suggested
by its bosonized form~(\ref{Jphi}), which converts this self-energy to a classical \emph{tree} graph in terms of $\vf$, just as occurs in
$D = 2$~\cite{BlaCabEM:2014}, then the torsional topological susceptibility $\vk$ will also be UV finite in $D = 4$. This interesting
possibility also merits an independent investigation.

Although one might expect the distance scale $1/\vk$ of non-trivial vacuum topology change to be of order $L_{\mathrm{Pl}}$,
and the value of $\vk$ to be of order of $M_{\mathrm{Pl}}$, there is no \emph{a priori} relation between the two scales. They are initially
distinct, just as $\La_{\mathrm{QCD}}$ and $f_\p$ are in QCD, to become possibly related only in a UV complete theory of quantum gravity.
Otherwise $\vk$ and $M_{\rm Pl}$ are treated as independent and unrelated dimensionful constants in the low energy EFT of gravity
proposed in this paper.

\end{document}